\documentclass[10pt,letterpaper]{article}
\usepackage[top=0.85in,left=2.75in,footskip=0.75in]{geometry}

\usepackage{amsmath,amssymb}

\usepackage{changepage}

\usepackage{textcomp,marvosym}

\usepackage{cite}

\usepackage{nameref,hyperref}

\usepackage[right]{lineno}

\usepackage[nopatch=eqnum]{microtype}
\DisableLigatures[f]{encoding = *, family = * }

\usepackage[table]{xcolor}

\usepackage{array}

\newcolumntype{+}{!{\vrule width 2pt}}

\newlength\savedwidth

\usepackage[aboveskip=1pt,labelfont=bf,labelsep=period,justification=raggedright,singlelinecheck=off]{caption}
\renewcommand{\figurename}{Fig}

\makeatletter
\renewcommand{\@biblabel}[1]{\quad#1.}
\makeatother

\usepackage{lastpage,fancyhdr,graphicx}
\usepackage{epstopdf}
\fancyheadoffset[L]{2.25in}
\fancyfootoffset[L]{2.25in}
\usepackage{graphicx}
\usepackage{url}
\usepackage{enumitem}
\usepackage{listings}
\usepackage{parcolumns}
\usepackage{xcolor}
\usepackage{booktabs}
\usepackage{caption}
\usepackage{verbatim}
\usepackage{subcaption}
\usepackage{amsmath}
\usepackage{array}
\usepackage{graphicx}
\usepackage{float}
\usepackage{tcolorbox}
\usepackage{hyperref}
\usepackage{tikz}
\usetikzlibrary{arrows.meta, positioning, shapes.geometric}

\usepackage[svgnames]{xcolor}
  \definecolor{diffstart}{named}{Grey}
  \definecolor{diffincl}{named}{Green}
  \definecolor{diffrem}{named}{OrangeRed}

\usepackage{listings}
  \lstdefinelanguage{diff}{
    basicstyle=\ttfamily\small,
    morecomment=[f][\color{diffstart}]{@@},
    morecomment=[f][\color{diffincl}]{+\ },
    morecomment=[f][\color{diffrem}]{-\ },
  }

\definecolor{codegreen}{rgb}{0,0.6,0}
\definecolor{codegray}{rgb}{0.5,0.5,0.5}
\definecolor{codepurple}{rgb}{0.58,0,0.82}
\definecolor{backcolour}{rgb}{0.95,0.95,0.92}

\lstdefinestyle{mystyle}{
    commentstyle=\color{codegreen},
    keywordstyle=\color{magenta},
    numberstyle=\tiny\color{codegray},
    stringstyle=\color{codepurple},
    basicstyle=\ttfamily\footnotesize,
    breakatwhitespace=false,         
    breaklines=true,                 
    captionpos=b,                    
    keepspaces=true,                 
    numbers=left,                    
    numbersep=5pt,                  
    showspaces=false,                
    showstringspaces=false,
    showtabs=false,                  
    tabsize=2
}

\newcommand\rev[1]{#1} 
 
\begin{document}
\vspace*{0.2in}

\begin{flushleft}
{\Large
\textbf{An empirical study of Java code improvements based on Stack Overflow answer edits} 
}
\newline
\\
In-on Wiratsin\textsuperscript{1,\Yinyang},
Chaiyong Ragkhitwetsagul\textsuperscript{1*,\Yinyang},
Matheus Paixao\textsuperscript{2},
Denis De Sousa\textsuperscript{2},
Pongpop Lapvikai\textsuperscript{1},
Peter Haddawy\textsuperscript{1}
\\
\bigskip
\textbf{1} Faculty of Information and Communication Technology, Mahidol University, Nakhon Pathom, Thailand
\\
\textbf{2} State University of Ceara (UECE), Fortaleza, Brazil
\\
\bigskip

%
%
\Yinyang These authors contributed equally to this work.





* chaiyong.rag@mahidol.ac.th

\end{flushleft}


%
\section*{Abstract}
Suboptimal code is prevalent in software systems.
Developers may write low-quality code for many reasons, including a lack of technical knowledge, insufficient experience, and time pressure, and once such code enters a codebase, its accumulation leads to significant maintenance costs and technical debt.

While programming, developers routinely consult external knowledge bases such as API documentation and Q\&A websites.
Stack Overflow is a popular Q\&A website where programmers ask questions and share code snippets. Under its community-curated content policy, its answers keep evolving as existing ones are edited.

In this paper, we present an empirical study of Stack Overflow Java answer edits and their application to improving code in open-source software projects.
We employ a modified code clone search tool to analyse Stack Overflow code snippets with version history, and apply it to open-source Java projects to \rev{identify code snippets that match earlier revisions of Stack Overflow answers and to surface the improvements subsequently applied to those answers}.
By analysing 874,438 Java accepted answers from SOTorrent, of which 140,840 had been edited, and 10,673 GitHub Java projects, we manually categorised the answer edits and created pull requests containing the code edits to open-source projects.
Our results show that \rev{16.11\%} of Stack Overflow Java accepted answers have more than one revision, with the average number of revisions per answer being 2.82. Moreover, \rev{49.31\%} of the code snippets in the Stack Overflow answer edits are considered applicable to open-source projects. \rev{A paired static-analysis comparison of the original and the latest snippets shows that the edits leave code structure and readability largely unchanged, as they are localised and semantic rather than restructuring, while tending to add defensive and error-handling logic.} \rev{Lastly, 19 out of the 80 code recommendations we submitted (23.75\%), including 11 out of 47 proposed bug fixes, were accepted by the GitHub project maintainers}.


%

\section{Introduction}

Suboptimal code is prevalent in software systems.  Developers may write low-quality code for various reasons, including a lack of technical knowledge~\cite{Shrestha2020}, limited experience~\cite{Eyolfson2011}, time pressure~\cite{Kuutila2020}, management decisions~\cite{Lavallee2015,Freire2020}, and even unhappiness~\cite{Graziotin2017}.
Regardless of the reason, lack of proper remediation may lead to large maintenance costs~\cite{Sjoberg2013} and technical debt~\cite{Tom2013,Bavota2016}.

Stack Overflow (SO)\footnote{\url{https://stackoverflow.com/}} is a popular website for programmers to ask questions and share their code snippets. 
The collaborative nature of SO has created a large source of programming knowledge.
As a result, programmers commonly reuse code from SO answers in their software projects~\cite{Yang2017,Ragkhitwetsagul2019,Zhang2019}. 
A recent study by Chen et al.~\cite{Chen2024} found that 9\% of SO answers have their code copied into open-source projects without modification. Other studies report that 6.32\% to 8.38\% of code in open-source projects is reused from SO~\cite{Huang2022,Chen2024b}.
\rev{Abdalkareem~\cite{Abdalkareem2025} confirms in a recent study that reusing crowdsourced code aids software development but may also degrade software quality through dependency overhead and increased maintenance effort.
Yu et al.~\cite{Yu2026} further characterise reusable code snippets by their prevalence, fault-resiliency, and longevity, providing criteria for prioritising high-quality reused snippets.}
With the launch of ChatGPT and other generative AI models, the popularity of SO is declining~\cite{Gordon2024,delRioChanona2024}. \rev{Xue et al.~\cite{Xue2026} report that the availability of large language models reduces SO question volume by 14--28\%, but the decline concentrates in routine, well-documented topics, resulting in niche partitioning rather than full displacement. Similarly, Shan et al.~\cite{Shan2025} find that generative AI changes, rather than eliminates, users' voluntary knowledge contribution on SO.} Indeed, a recent study~\cite{Helic2025} shows that SO remains useful for developers, as the questions and answers on SO are evolving to address complex and challenging problems that are not handled by such generative AIs.

Although code snippets from SO may contain issues such as security vulnerabilities~\cite{Acar2016}, API misuses~\cite{Zhang2018}, and license-violating code~\cite{Ragkhitwetsagul2019,Zhang2019}, the crowdsourcing nature of the website allows for constant community updates that not only optimise the answers' code but also fix potential issues~\cite{Diamantopoulos2019,Tang2021}.
A recent study~\cite{Sheikhaei2023} shows that 55.3\% of update request comments on SO are addressed within 24 hours. Moreover, Mondal et al.~\cite{Mondal2025} found that 53.3\% of the answer edits improve the answers to match the questions, and 51\% of the edits aim to optimise code performance.
\rev{In a similar vein, Zuo et al.~\cite{Zuo2025} track how security-relevant C\# snippets evolve through user revisions, cautioning that the quality of edits depend on the editor's experience.
Jallow et al.~\cite{Jallow2025} further show that SO code snippets are non-stationary over time and recommend treating SO as a time-series data source, which reinforces the value of analysing the version histories of answer edits.}

Several studies have proposed approaches that leverage SO knowledge to assist developers. 
Some of them create automated tools that provide working code examples~\cite{Keivanloo2014}, show relevant SO posts based on the code context in the IDE~\cite{Ponzanelli2013,Ponzanelli2014}, and improve API documentation~\cite{Treude2016}. 
Tang and Nadi~\cite{Tang2021} proposed a method to automatically identify comment-edit pairs on SO, i.e., comments in an answer that trigger an edit in the answer, and recommend the edits to projects with the same outdated snippet.
Gao et al.~\cite{Gao2023} propose a technique to retrieve semantically equivalent questions from SO and provide a ranked list of code snippets as recommendations.
Jallow et al.~\cite{Jallow2024} examine the adoption of outdated code snippets from SO, focusing on those that contain security issues. They reported 43 popular GitHub projects missing fixes to bugs and
vulnerabilities based on the answers on SO.

Nonetheless, a gap remains in these existing studies.
First, the recommendations are limited to the code context in the IDE.
Second, matches are only made to the latest version of code answers on SO, so that possible matches to earlier versions are missed.  
Third, the studies focusing on Stack Overflow updates are based on updates triggered by comments, excluding answer edits for other reasons. 
Fourth, the recommendations were based solely on exact matches between code on SO and in software projects, overlooking code that had been modified. Lastly, they studied only popular projects on GitHub, ignoring the less popular ones that might also benefit from SO code recommendations.

In this paper, we fill the gap by performing an empirical study of SO Java answer edits and their recommendations as potential code improvements. We analysed 874,438 Java accepted answers on SO and identified the 140,840 of them that had been edited after they were posted. 
Leveraging a scalable code clone search tool with optimised configurations for locating code clones, with exact matches and with modifications, we located matching code snippets between the collected SO accepted answers and 10,673 Java open-source projects. The projects were classified into three tiers based on their popularity.
By comparing \rev{the matched project code with the newer version of the corresponding answer} (i.e., the proposed update) on SO, we classified the proposed updates into groups based on their update types.
Lastly, we investigated whether developers found the proposed updates based on SO answer edits useful by opening pull requests containing those edits.
\rev{In contrast to prior work, which recommends only the latest version of an answer, relies on comment-triggered edits, matches code exactly, or studies only popular projects, our study is the first to jointly (1) drive recommendations from the full answer edit history, matching project code against earlier revisions; (2) consider all answer edits regardless of what triggered them; (3) match code with modifications rather than exact clones only; and (4) validate the recommendations end-to-end through real pull requests across projects of all popularity levels.}

This paper makes the following contributions.
\begin{enumerate}
    \item A large-scale study of SO answer edits analysing the full edit history of the 140,840 edited Java accepted answers and 10,673 Java open-source projects covering both less-popular and highly-popular projects.
    \item A specialised clone search tool that supports code revisions on SO, which is made publicly available\footnote{\url{https://github.com/cragkhit/Matcha}}.
    \item A classification of code improvements based on the Stack Overflow answer edits.
    \item A dataset of manually validated code clones between SO and GitHub projects that are made publicly available, which can be used for future clone studies.
\end{enumerate}

This empirical study provides insights into the categories of SO Java answer edits and their applications to open-source software projects, and provides a basis for automated recommendations for code updates based on SO answer edits.

\subsection{Motivating example}
\label{subsec:motivating_example}

As motivation for this study, we take a look at the SO question number 40665315\footnote{\url{https://stackoverflow.com/questions/40665315}}.
In the question, the developer is using the REST Assured\footnote{\url{https://rest-assured.io/}} library to write unit tests for a REST application developed with the Spring Boot\footnote{\url{https://spring.io/projects/spring-boot}} framework. 
For this particular unit test, the developer needs to specify the port on which the service will run, which is the main topic of the question.
A solution was given and was marked as the accepted answer in November 2016.

\begin{figure}[H]
\centering
    \includegraphics[width=0.7\linewidth]{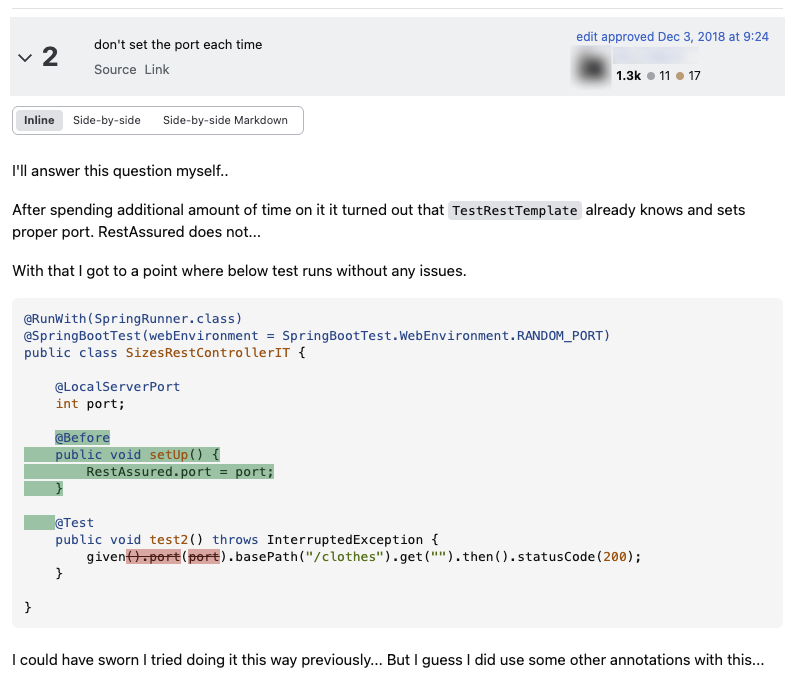}
\caption{{\bf An example of a code answer edit on a Stack Overflow accepted answer.} The edit optimises the code by moving the port setup to the \texttt{@Before} method.}
\label{fig:motivating_example}
\end{figure}

However, in December 2018, the accepted answer was edited further as shown in Fig~\ref{fig:motivating_example}.
We can see that the new \texttt{setUp()} method has been added, and the call to method \texttt{.port()} in the test has been removed. This update eliminates the need to set up the port every time the test is executed, thereby saving resources.

We conducted an initial search for instances of the outdated code-answer version in this answer on GitHub and found that 7 out of 10 projects are still using this sub-optimal solution.
As this example shows, suboptimal solutions can be posted (and accepted) on SO, and it may take the community a while to optimise them. 
In the meantime, developers may use the sub-optimal snippets in their own solutions.
This example illustrates that software projects that use suboptimal code may benefit from adopting the more recent, optimal code solutions on SO.


\subsection{Research questions}

\rev{Our study is designed as a pipeline that moves from establishing the supply of Stack Overflow answer edits, to matching those edits against source code in GitHub projects, and finally to validating their usefulness in practice. We first quantify how often accepted Java answers are edited after they are posted, which establishes the pool of potential code improvements available for recommendation (RQ1). We then examine, for code found in open-source projects that is similar to an earlier revision of such an answer, what types of improvements the corresponding answer edits suggest and how these are distributed across projects of differing popularity (RQ2). Because that classification is manual and therefore subjective, we then measure the same edits objectively, quantifying how they alter the structure and readability of the code (RQ3). Finally, to confirm that these matched edits translate into changes that developers actually value, we submit a subset of them as pull requests to the corresponding projects (Section~\ref{sec:results}, Phase~4). This last step is a practical validation of the pipeline rather than a separately declared research question.}

This study seeks to answer the following research questions:
\begin{itemize}

\item \textbf{RQ1:} \textit{\rev{To what extent are accepted Java answers on Stack Overflow edited?}} For this RQ, we want to understand the prevalence of edits to SO Java accepted answers. We extract the revision history of SO Java accepted answers to determine how many answers were edited after they were posted and how large those edits are. \rev{This establishes the supply of potential code improvements that could later be recommended to developers.}

\item \textbf{RQ2:} \textit{\rev{What types of code improvements do Stack Overflow answer edits suggest for code found in open-source projects, and how are they distributed across projects of different popularity?}}
\rev{For this RQ, we use the supply of potential code improvements from RQ1 to (1) gain deeper understanding of the types of code improvements suggested by SO answer edits, and (2) determine how often these improvements are applicable to open-source projects. We classify the answer edits into categories based on the types of code improvements they suggest, and we examine how these categories are distributed across projects of differing popularity.}

\item \rev{\textbf{RQ3:} \textit{What are the differences in terms of code structure and readability introduced by Stack Overflow answer edits?}}
\rev{For this RQ, we objectively measure the differences between the earlier and later revisions of SO answer edits in terms of code structure and readability. This provides a quantitative assessment of the impact of the answer edits on the code, complementing the manual classification in RQ2.}

\end{itemize}

\rev{To scope our research questions, our approach matches project code against earlier revisions of SO answers using a code clone search tool, which establishes textual similarity but not the origin or the direction of the reuse. RQ2 therefore asks what improvements the corresponding answer edits suggest for such code and whether they are applicable to it, and the pull-request evaluation asks whether maintainers find those improvements useful. Neither question measures how much code in open-source projects was actually copied from SO, nor the causal impact of SO on those projects. We return to this assumption, and to the alternative explanations for a matched pair, in the construct validity discussion in Section~\ref{sec:threats}.}



\section{Related work} 
\label{sec:background}

\subsection{Stack Overflow answer edits}
A few previous studies have examined aspects of SO answer edits. 
Wang et al.~\cite{Wang2020} investigate how the SO badge system, which rewards users based on the quantity of revisions, incentivised answer revision behaviour and quality. Analysing over 2,377,692 SO answers, the study reveals that users make significantly more revisions on badge-awarding days than on normal days. These revision spikes are associated with a higher likelihood of the revision being rolled back (i.e., rejected), often due to undesired code/text formatting or incorrect changes. During these spikes, users tend to make small, text-only revisions (e.g., text correction and code formatting) rather than complex code corrections or improvements.
Baltes et al.~\cite{Baltes2020} similarly analysed the SO post edits, but based on their edit comments. They reported that 17.5\% modified only a code block, 64.2\% only a text block, and
18.3\% both text and code blocks of a particular SO post. They also categorised the post edits into 25 categories (e.g., Formatting, Code, Fixing, Improving) by manually analysing frequent messages and using regular expressions. Formatting, adding, and fixing were the most frequent categories across all edits, as well as in the subset that modified code blocks only.
Specifically addressing non-functional properties, the authors identified a few thousand code edits that mentioned performance, size, and memory. The findings suggest that SO data can serve as a treasure trove for future work on mining fine-grained patches. 

Tang and Nadi~\cite{Tang2021} follow the same line of research as they investigate the potential of using SO comment-edit pairs as a complementary data source for building code maintenance data sets, which are crucial for tools like program repair and code recommenders. The authors propose an automated technique to extract these pairs, matching a comment suggesting a change to a subsequent code edit and providing concrete explanations for the change. Applying their technique to five popular programming languages (Java, JavaScript, Android, PHP, Python), they find that the extracted edits are rarely tangled (11\%) and that 27\% of the confirmed pairs are potentially useful for code maintenance applications. The study resulted in 15 pull requests to popular open-source repositories based on the mined pairs, 10 of which were accepted.
Sheikhaei et al.~\cite{Sheikhaei2023} study the update request comments (URCs), i.e., comments on SO that point out issues like buggy code or obsolete information, that trigger an update to the answer post and improve its quality. Through an empirical analysis of 1,221 comments, they find that URCs are prevalent (51.7\%) and that, while 55.3\% of URCs are addressed within 24 hours, 36.5\% remain unaddressed after a year and the community voting mechanism is ineffective at surfacing these issues. 

Zhang et al.~\cite{Zhang2022} focus specifically on C/C++-related answers and revisions on SO and their security weaknesses.
They found that a significant 36\% of security weaknesses (CWEs) were detected, and while revisions generally reduce these weaknesses, a majority of vulnerable snippets are never corrected, exposing users to risks.
Jallow et al.~\cite{Jallow2024} found that 51\% of SO answers in open-source projects become outdated because developers fail to track the original source, leading to missed updates, including fixes for up to 15 distinct security issues in 43 popular GitHub projects. For C/C++ code, this includes prevalent flaws such as buffer overflows. They suggest that tools are urgently needed to monitor SO for fixes and effectively surface the best, most current code solutions.

Mondal et al.~\cite{Mondal2025} investigate whether SO answer edits enhance the answer quality. Analysing 94,994 edited Python answers on SO, they compare the initial and latest versions across six key dimensions: semantic relevance, code usability, complexity, security, optimisation, and readability. The findings show that the answer edits offer both benefits and drawbacks. The benefits include improving semantic relevance in 53.3\% of cases and making 9\% of previously broken code executable. On the other hand, there are negative impacts, such as increased code complexity (32.3\%), decreased readability (49.7\%), and introducing new security vulnerabilities in 20.5\% of cases rather than fixing existing ones.

\subsection{Generating code recommendations}
Previous work has explored basic approaches to generating code recommendations from SO data. 
The work of Ponzanelli et al.~\cite{Ponzanelli2013,Ponzanelli2014} presents \textsc{Prompter}, an IDE plugin that recommends Stack Overflow discussions based on the developer's current coding context.
By leveraging existing search engines, \textsc{Prompter} identifies relevant discussions based on a predefined threshold.
A developer working in an IDE can open and visualise the Stack Overflow discussion directly in the IDE.
Although it does not directly recommend code snippets to developers, \textsc{Prompter} leverages Stack Overflow's knowledge to enhance the developer experience in the IDE by providing more context for their coding decisions.
On a similar note, the work by Rahman et al.~\cite{Rahman2016} proposes \textsc{RACK}, an automated tool that focuses on API recommendations based on Stack Overflow knowledge.
Unlike the previously mentioned work, the \textsc{RACK} tool utilises natural language queries to search for relevant Stack Overflow posts. The work by Rubei et al.~\cite{Rubei2020} aims to recommend relevant posts based on a developer's coding context.
The proposed tool, called \textsc{PostFinder}, focuses on enhancing both Stack Overflow posts and developers' code with additional metadata to improve matching.
The results demonstrate that the new features facilitate matching code context with highly relevant posts.
Gao et al.~\cite{Gao2023} propose QUE2CODE, a two-stage, query-driven model consisting of semantically equivalent question retrieval and best code snippet recommendation, designed to recommend the best code snippets from Stack Overflow posts to address developer challenges of query mismatch and information overload. 

The community working on code recommendations has been rapidly growing, with papers proposing creative ways to recommend code from sources other than Stack Overflow.
Keivanloo et al.~\cite{Keivanloo2014} propose an approach to find working code examples from a large code corpus on the Internet by using maximal frequent itemset mining and a custom search ranking function. 
Nyamawe et al.~\cite{Nyamawe2018} recommend refactoring solutions by using traceability and code metrics. 
\textsc{Coder}~\cite{Jin2023} is a graph-based code recommendation framework that considers both user-code and user-project interactions to enhance recommendations.
Aroma~\cite{Luan2019} is a tool that provides code recommendations via a structural code search. 
Guo et al.~\cite{Guo2024} explore the potential of ChatGPT in refining existing code based on human-written code reviews.

\section{Materials and methods}
\label{sec:methodology}
This section describes the methodology we employed to answer each of our \rev{three} research questions, including the dataset, the code clone search tool, and the experimental framework.

\subsection{Datasets} 
\label{subsec:sotorrent}
The SO data for this study were retrieved from the SOTorrent~\cite{Baltes2018a} dataset, the largest SO dataset to date. The dataset is created from the SO data dump, augmented with the version history of SO content. The version history can be retrieved at the level of the whole post or individual post block. The SOTorrent dataset has been widely used by several previous studies~\cite{Baltes2020,Manes2019,Bangash2019,Nugroho2022}.
In its latest version (version 2020-12-31), the dataset contains the content of 51,296,931 SO posts with 81,536,422 post versions.
The dataset can be accessed via Google BigQuery\footnote{\url{https://console.cloud.google.com/bigquery?project=sotorrent-org}} or Zenodo\footnote{\url{https://zenodo.org/record/3746061}}.

\rev{\paragraph{Stack Overflow code snippets and their revisions.} 
We worked with a local MySQL copy of the SOTorrent 2020-12-31 release.
Because tags in Stack Overflow are attached to questions rather than answers, Java answers were identified through their parent question, by joining \texttt{Posts} to \texttt{PostTags} on the question and resolving the \texttt{java} tag to its \texttt{Tags.Id}.
An answer entered our set when it was both \emph{accepted}, i.e., its \texttt{Id} equals the parent question's \texttt{AcceptedAnswerId}, and \emph{revised}, i.e., it has at least one \texttt{PostVersion} row whose \texttt{PredPostHistoryId} is not null, which marks a version that has a predecessor and therefore an edit.
For each of those answers we extracted the code blocks from \texttt{PostBlockVersion}, in which code blocks are distinguished from prose blocks by \texttt{PostBlockTypeId = 2}.
Answers containing more than one code block were neither merged nor reduced to a single snippet.
Each block is instead tracked separately by its \texttt{LocalId}, which gives the block's position within a revision and remains stable across the revisions in which that block exists, so an answer with two code blocks contributes two independent edit histories.
For every \texttt{(PostId, LocalId)} pair we wrote out each revision of the block as its own file, along with two distinguished versions: the \emph{latest} version, i.e., the row with \texttt{MostRecentVersion = 1}, and the \emph{original} version, i.e., the root of the block's edit chain given by \texttt{RootPostBlockVersionId}.
We take the original from the root of the chain rather than from the answer's first revision because an edit chain can break and restart when a block is removed and later re-added; in that case the earliest ancestor of the current chain, rather than the very first version of the post, is the snippet from which the latest version was actually derived.
}

\paragraph{GitHub open-source Java projects}
\rev{To collect} GitHub projects, we employed GHS (GitHub Search)~\cite{Dabic:msr2021data}, a tool and dataset tailored for mining software repositories, specifically for searching GitHub projects.
The GHS dataset comprises 735,669 repositories written in 10 programming languages.
For each project, GHS provides data on 25 characteristics that can be used as filters in the search.
For this exploratory study, we searched GHS with the following filters: \texttt{Language: Java; Exclude Forks; Has Open Issues; Has Open Pull Requests.} 
We only selected projects written in \texttt{Java} because it is one of the most widely used programming languages~\cite{IEEETopProg2024} and is supported by our selected code clone search tool (see Section~\ref{subsec:siamese}).
We excluded all forks to ensure our dataset contained no redundant projects, and selected projects with both open issues and open pull requests to ensure their aliveness.

\subsection{Code clone search tool}
\label{subsec:siamese}
To locate \rev{code in open-source projects that is similar to Stack Overflow code snippets}, we chose Siamese~\cite{Ragkhitwetsagul2019} for this study. Siamese is a code clone search tool that is both accurate and scalable, capable of handling hundreds of millions of lines of code. The tool encompasses multiple code representations by transforming code into various forms to capture different types of code clones, query reduction that retains only highly relevant terms in the code query, and a customised ranking function that enables the selection of a preferred clone type to be ranked first. 
It offers an accurate clone search with high precision and recall for Type-1 (i.e., code clone with exact copies) to Type-3 clones (i.e., code clones with modified/added/removed statements), outperforming other state-of-the-art code search and code clone detection tools~\cite{Zakeri2023}. Although numerous code clone detection and search tools exist, only a few are scalable to large-scale code bases with millions of lines of code. A scalability comparison with six existing tools reveals that only Siamese and SourcererCC~\cite{Sajnani2016} can scale to a code corpus of 365 million lines of code; however, Siamese can return clone search results in just 8 seconds.




\subsection{Experimental design}
\begin{figure}[H]
\includegraphics[width=\textwidth]{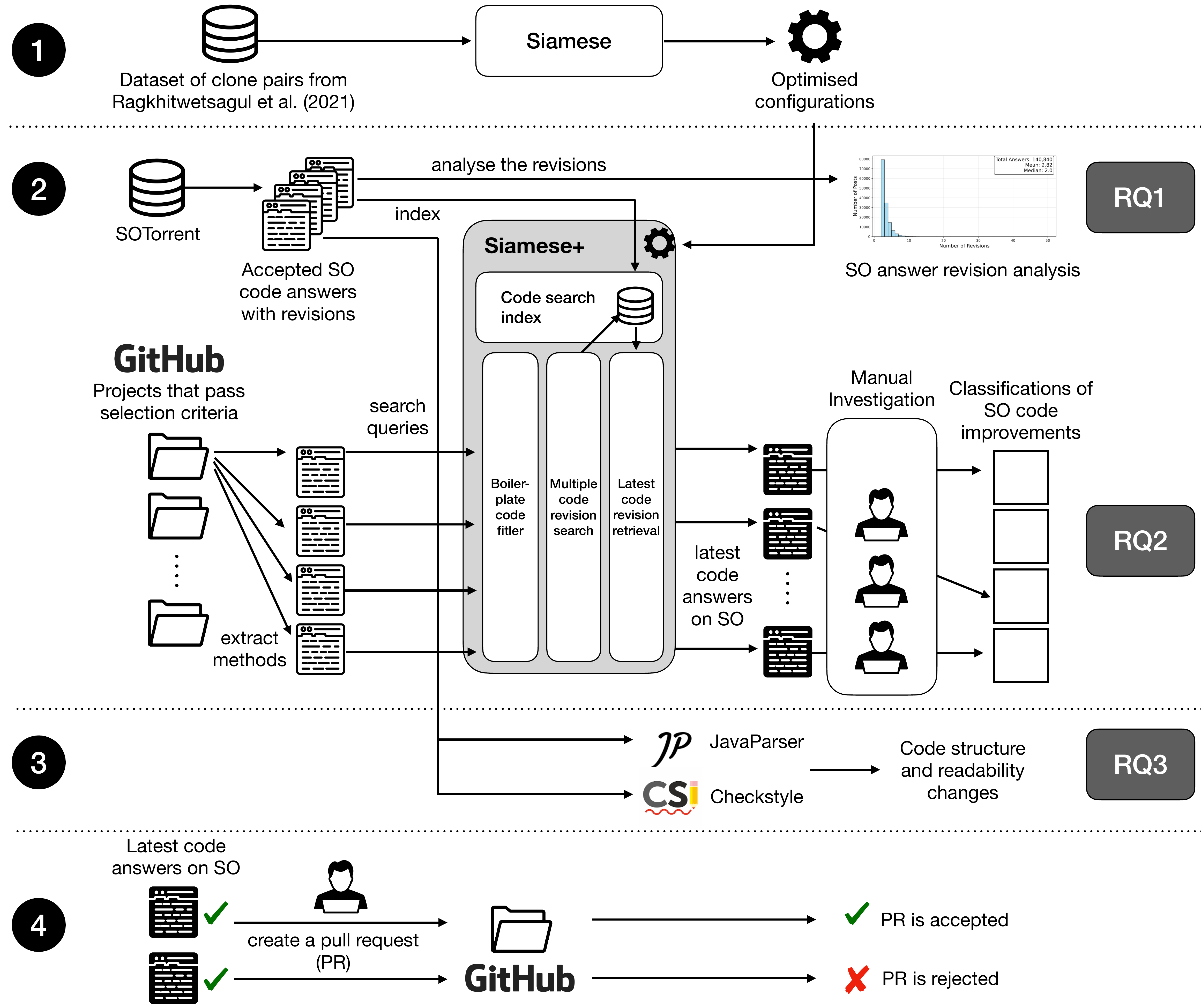}
\caption{\rev{{\bf Overview of our four-phase exploratory study.} Phase~1 tunes the Siamese clone search tool. Phase~2 detects clones between Stack Overflow answer revisions and GitHub projects (answering RQ1 and RQ2). Phase~3 compares the original and the latest version of the Stack Overflow answers with two static analysers (answering RQ3). In Phase~4, the answer edits are submitted as pull requests.}}
\label{fig:overview}
\end{figure}

Fig~\ref{fig:overview} shows the experimental design of our study.
The study was separated into four sequential phases, described in the subsections below.
All the code and data necessary to replicate this study are available in our replication package\footnote{\url{https://doi.org/10.5281/zenodo.17220745}}.
This section provides a detailed description of each phase of our study.

\subsubsection{Phase 1: Optimising configuration and modifying the code clone search tool}
\label{subsec:phase1} 

This phase is a preparation for the experiment. We first used a search-based technique to determine the optimised configurations of Siamese. We then augmented Siamese to detect SO's outdated code in software projects and enable it to provide the latest code answer based on answer edits. We explain these steps in detail below. We call this modified version ``Siamese+'' and use it for the subsequent phases of the study.

\paragraph{Step 1 Optimising Siamese configurations:}
Because the accuracy of code similarity analysers depends heavily on their configuration, and default settings are not always optimal~\cite{Ragkhitwetsagul2018}, our first step is to tune Siamese for locating similar code across SO answers and Java projects.
This tuning is especially important because Siamese has not previously been evaluated on SO--GitHub matching, and such clones can differ from regular clones (e.g., incomplete snippets and more boilerplate code~\cite{Ragkhitwetsagul2019a}).



To perform the configuration optimisation, we leveraged the ground truth dataset\footnote{\url{https://ucl-crest.github.io/cloverflow-web/downloads.html}} introduced by a previous study~\cite{Ragkhitwetsagul2021}.
This dataset contains 2,302 pairs of online code clones identified between SO and Java projects in the Qualitas corpus~\cite{tempero2010qualitas}, a curated collection of 111 Java projects spanning across different domains.
These clone pairs were initially detected by SourcererCC~\cite{Sajnani2016} and Simian\footnote{\url{https://simian.quandarypeak.com/}}, and then manually reviewed by the authors.
The identified clones were categorised into seven patterns of code cloning. For this study, we retained only the three patterns representing code reusable between real Java projects and Stack Overflow: Cloned from Qualitas project to Stack Overflow (QS), Cloned from an external source to Stack Overflow (EX), and Cloned from each other or from an external source outside the project (UD); the remaining four patterns (e.g., boilerplate and not-clones) were excluded.
This refinement resulted in a final dataset comprising 553 clone pairs.

The parameter tuning approach focused on maximising the mean reciprocal rank (MRR) metric.  
MRR was chosen because it aligns with the study's goal: finding the best single match between SO answers and code in Java software projects.
The MRR metric is defined as shown in Equation~\ref{eq:mrr}, where $Q$ represents the set of all code snippets (queries) in the ground truth, each snippet is denoted as $q \in Q$, and the Reciprocal Rank (RR) is calculated using Equation~\ref{eq:rr}.

\begin{equation}
MRR(Q) = \frac{\sum_{q=1}^{|Q|} RR(q)}{|Q|}
\label{eq:mrr}
\end{equation}

\begin{equation}
\begin{aligned}
\mathit{RR}(q) & = \frac{1}{rank(S(q))}, & \text{if $|S(q)| > 0$} \\
\mathit{RR}(q) & = 0, & \text{if $|S(q)| = 0$}
\end{aligned}
\label{eq:rr}
\end{equation}

To assess Siamese using the ground truth, we indexed the code snippets extracted from the Java projects in the Qualitas corpus and performed searches using code snippets from SO. 
The SO code snippets were obtained from the StackExchange archive\footnote{\url{https://archive.org/details/stackexchange}}; for each ground-truth post, we extracted the relevant snippet by its recorded start and end lines, treating posts with multiple clone-pair snippets individually.

The Java projects from the Qualitas benchmark were obtained from the official website\footnote{\url{http://qualitascorpus.com/download/}}. 
For this study, we retrieved the release version 20130901r, which matches the one used in constructing the ground truth~\cite{Ragkhitwetsagul2021}. 
In total, the Qualitas benchmark used in this empirical analysis comprises 111 projects, encompassing 166,709 Java files and 19,614,083 lines of code. 
The only preprocessing operations applied to the Qualitas source code involved the removal of code comments and the use of pretty printing\footnote{\url{https://astyle.sourceforge.net}} to standardise the code structure. 

We employed Grid Search as our search algorithm to optimise the Siamese configurations.
Despite its simplicity, it avoids the common pitfalls reported in the literature regarding clone detector tuning~\cite{wang2013searching,Ragkhitwetsagul2016}. \rev{Being exhaustive, it is also deterministic and reproducible, and it guarantees that the configuration we report is optimal with respect to the grid, which a stochastic search such as Random or Bayesian search cannot.}
Table~\ref{tab:siamese_params} lists the Siamese parameters we considered.
We selected the same parameters used in the original Siamese publication~\cite{Ragkhitwetsagul2019} and expanded by considering a larger set of values for each parameter. Then, we selected a pre-defined set of values in the search space (a grid).
The chosen values included the n-gram size values of \{4, 6, 8\}, \rev{a QR threshold drawn from \{8, 10\} and a boosting value drawn from \{-1, 10\} for each of the four code representations}, minCloneSize values of \{6, 10\} and two sets of simThreshold values of \{20\%,40\%,60\%,80\%\} and \{30\%,50\%,70\%,90\%\}.
The Grid Search algorithm then performed an exhaustive search over the grid-defined search space\rev{, i.e., $3 \times 2^{10} = 3,072$}.
\rev{The grid samples the value ranges in Table~\ref{tab:siamese_params} rather than enumerating them in full, because a complete enumeration is computationally infeasible: even the sampled grid required 3,072 configurations and approximately 54 hours to evaluate (Section~\ref{sec:results}).
We also note that this tuning is performed entirely on data that is disjoint from the data analysed in the study itself, i.e., on the 111 Qualitas projects and the ground-truth SO snippets.}

\begin{table}
    \footnotesize
    \caption{Siamese parameters considered in this study.}
    \begin{tabular}{lp{9cm}}
        \toprule
        \textbf{Parameter} & \textbf{Values}\\
        \midrule
        \textbf{n-gram size} & 4, 5, 6, 7, 8, 9, 10, 11, 12, 13, 14, 15, 16, 17, 18, 19, 20, 21, 22, 23, 24\\
        \textbf{QR thresholds} & 2, 4, 6, 8, 10, 12, 14, 16, 18, 20 \\
        \textbf{boosting } & -1, 1, 2, 4, 6, 8, 10, 12, 14, 16, 18, 20 \\
        \textbf{minCloneSize} & 6, 7, 8, 9, 10, 11, 12, 13, 14, 15, 16 \\
        \textbf{simThreshold} & \{10\%,20\%,30\%,40\%\}, \{20\%,30\%,40\%,50\%\}, \{30\%,40\%,50\%,60\%\}, \{40\%,50\%,60\%,70\%\}, \{50\%,60\%,70\%,80\%\}, \{60\%,70\%,80\%,90\%\}, \{10\%,30\%,40\%,50\%\},
        \{20\%,40\%,60\%,70\%\}, \{30\%,60\%,80\%,90\%\}, \rev{\{20\%,40\%,60\%,80\%\}, \{30\%,50\%,70\%,90\%\}} \\
        \bottomrule
    \end{tabular}
    \label{tab:siamese_params}
    \vspace{-0.5cm}
\end{table}



\paragraph{Step 2 Modifications of the code clone search tool:}
To support the goal of this study, we augmented Siamese to include three additional modules: (1) boiler-plate code filter, (2) multiple-code revision search, and (3) latest code revision retrieval (as shown in Fig~\ref{fig:overview}). 
The three modules are crucial for analysing code snippets in the revisions of Stack Overflow answers and locating the latest version. 
Although the input is provided to Siamese as a project, the tool reads each Java file in the project, parses it, and performs the code clone search at the method level, one at a time. 
Within the search process, we incorporated the three additional modules to enable Siamese to return code recommendations as follows.

\textbf{Boilerplate Code Filter:}~
Clones between Stack Overflow and software projects often contain boilerplate snippets (e.g., \texttt{getter}, \texttt{setter}, \texttt{equals()}, \texttt{compareTo()}, \texttt{toString()} methods) that are not useful to developers. Following the boilerplate classification of a previous study~\cite{Ragkhitwetsagul2019a}, we encoded such patterns as a set of regular expressions and integrated them into Siamese as a query filter. Any code query matching the filter is excluded from the search, thereby avoiding trivial recommendations.

\textbf{Multiple Code Revision Search:}~
This module lets Siamese search across all revisions of a code snippet in an SO answer. To enable it, we built a Siamese clone search index containing every revision of each Java accepted answer.
We named the code snippet in each revision by concatenating its \texttt{PostId}, \texttt{LocalId} (the code block within the answer), and \texttt{HistoryId} (the revision), as defined by SOTorrent, marking the first and last revisions as \texttt{original} and \texttt{latest} (e.g., \texttt{8394534\_0\_original.java} and \texttt{8394534\_0\_latest.java}). This lets us determine from the first-ranked Siamese result whether the matched snippet is the latest version by inspecting its file name.

\textbf{Latest Code Revision Retrieval:}~
This module allows Siamese to return the latest revision of the matched code snippet. Based on the results from the multiple code revision search module, Siamese checks whether the latest version of the SO answer is returned for each code snippet. If not, Siamese includes the latest version of the answer in the list of code recommendations. 
After completing the search using all the methods in the given Java project, Siamese returns a list of code recommendations in CSV format. Each record contains the file name, method name, start line, and end line of code in the project, as well as the \texttt{PostId} of the Stack Overflow post containing the latest code revision.


\subsubsection{Phase 2: Detecting clones between Stack Overflow and GitHub}
\label{subsec:phase2}

In Phase 2, we conducted a code clone search across SO's Java accepted answers and GitHub's Java projects.

\paragraph{Indexing Stack Overflow Java accepted answers with all revisions:}
Our analysis of Java posts on SOTorrent data shows that there are \rev{2,832,800 Java answers on Stack Overflow, of which 874,438 are accepted answers}. From these accepted answers, there were 140,840 Java accepted answers that had at least one revision.
First, we extracted code snippets with revisions from all accepted Java answers. 
Next, we extracted 283,838 code snippets of all the revisions of the Java-accepted answers and imported them into the Siamese+ clone search index. 
This indexing phase occurs only once in the study, and subsequent use of Siamese+ will be via a query to a similar code snippet, which takes significantly less time (e.g., seconds).
\paragraph{Filtering of GitHub projects:}
\label{sec:filter}
As noted earlier, we assume that SO answer edits may provide optimised snippets that improve project codebases, and that projects of different popularity, used here as a proxy for code quality, may benefit differently: higher-popularity codebases likely contain fewer sub-optimal snippets and yield fewer recommendations than lower-popularity ones. To evaluate this assumption, we needed a representative sample of projects spanning different code-quality levels.

For each project, GHS provides 3 popularity metrics: \texttt{Number of Stars}, \texttt{Number of Watchers}, and \texttt{Number of Forks}.
GitHub popularity metrics, such as those provided by GHS, have been used as quality proxies for GitHub projects~\cite{Zampetti2019,Gonzalez2020,Sheoran2014}.
Although isolated popularity metrics, such as stars, have been shown not to be the most effective way to assess a project's code quality~\cite{Munaiah2017}, we believe that combining these metrics could yield a trustworthy proxy.
Hence, for each popularity metric provided by GHS, we computed the distribution and divided it into quartiles. 
Projects that appeared above the third quartile for all metrics were considered to have a high-popularity codebase.
Similarly, projects that appeared below the first quartile for all metrics were considered to have a low-popularity codebase.
Finally, a project that fell between the first and third quartiles for all metrics was considered to have a medium-popularity codebase.
Projects that appear in the intersection of quartiles between metrics, e.g., above the third quartile for a metric and below the first quartile for another metric, were excluded from this study.
These criteria provided a clear separation of projects based on the proxy we employed for the popularity metrics.
\rev{The rule is deliberately conservative. By requiring the three metrics to agree before a project is assigned to the high- or the low-popularity group, and by discarding the projects that give conflicting signals rather than forcing them into a group, it favours a clean separation between the groups over retaining the largest possible sample. We consider this trade-off worthwhile because the comparison across the three popularity groups is central to RQ2, and a group boundary blurred by ambiguous projects would weaken that comparison more than the reduction in sample size does.}

\begin{figure}[H]
      \includegraphics[width=0.8\columnwidth]{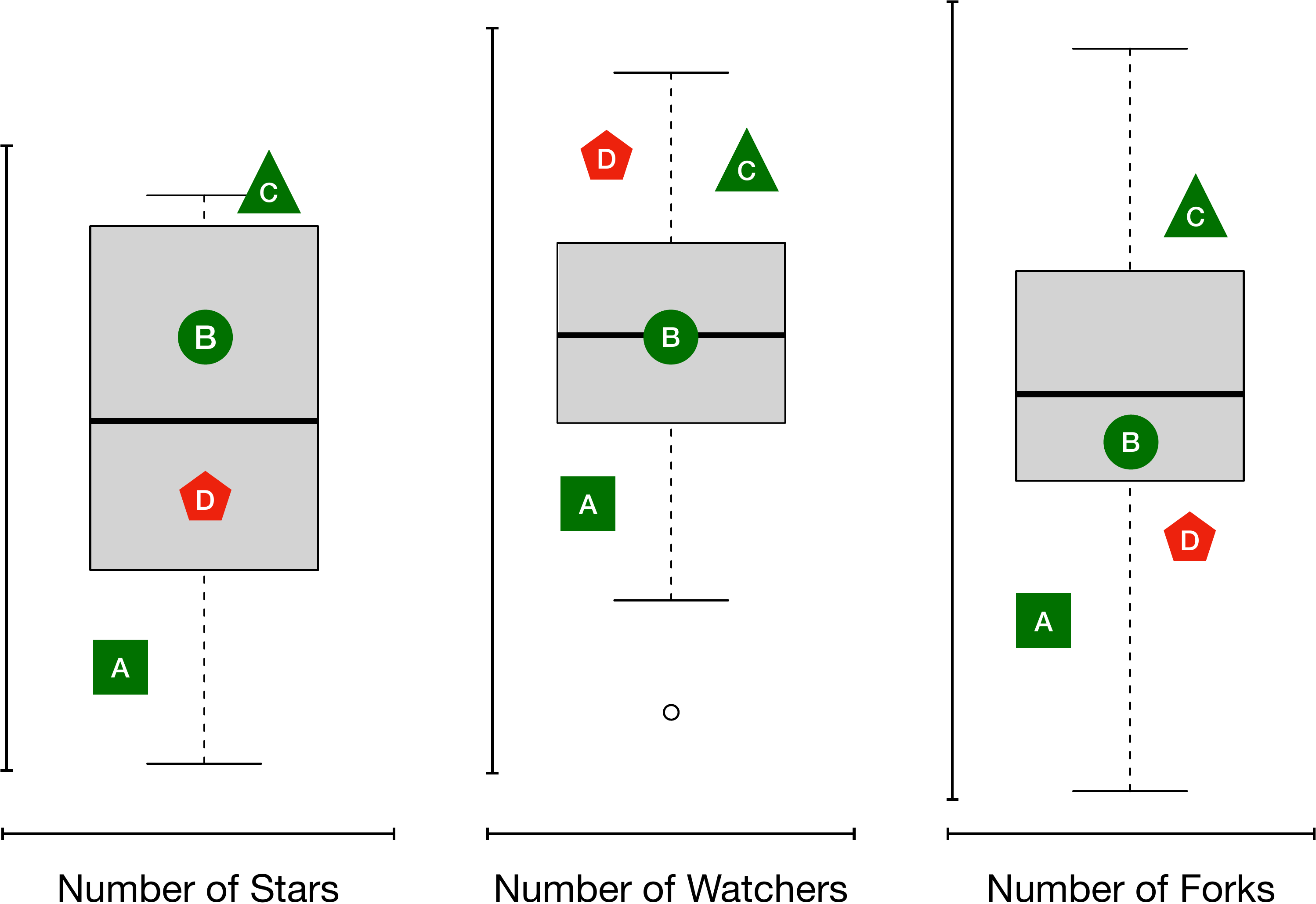}
\caption{{\bf GitHub project selection criteria.} The criteria are based on the distributions of the number of stars, watchers, and forks.}
\label{fig:boxplots}
\end{figure}

Fig~\ref{fig:boxplots} depicts our project selection criteria. 
For example, among four projects (A--D) whose star, watcher, and fork distributions are shown, Projects~A, B, and~C are included because they fall consistently below the first quartile, between the quartiles, and above the third quartile across all three metrics, respectively, whereas Project~D is excluded because it lands in different quartiles for different metrics.

\paragraph{Searching for clones over revisions:}
After selecting the projects, we ran Siamese+ for each project.
First, for a given software project, Siamese+ extracts the code for all the project's methods and applies the boilerplate code filter. 
Next, for each remaining method, Siamese+ searches its index, which contains the code for SO's accepted answers, including all revisions.
If Siamese+ locates a snippet from an older revision of a Stack Overflow answer (i.e., not the latest), it returns the latest version of the answer as a recommendation.

\paragraph{Manual classifications of potential clones:}
After running Siamese+ for all selected projects and collecting all recommendations, we entered the final step of Phase 2.
\rev{Before describing the classification procedure, we clarify the criteria underpinning it. We considered a GitHub code snippet to be \emph{outdated} if it matched an earlier revision of an accepted SO answer and that answer was subsequently updated with changes to its code. We did not independently judge whether the earlier code was objectively incorrect or unoptimised. Instead, we treated the latest revision of the accepted answer as the reference version, since it represents the most up-to-date solution maintained through the collaborative editing process of the SO community. The differences between the earlier and the latest revisions were then analysed and classified into improvement categories, as described below and reported in Section~\ref{subsubsec:classification}.}
For each potential clone, we looked at the original SO answer, the latest SO answer, the GitHub project code and performed a manual classification. 
\rev{First, we manually checked each recommended Stack Overflow-GitHub pair to determine whether they were code clones, i.e., similar code with some level of similarity, to ensure the correctness of the recommendations by Siamese+. Next, we continued with the categorisation of the recommendations.}
We initially adopted the SO post-edit categorisation proposed by Baltes et al.~\cite{Baltes2020} (see Table~\ref{tab:baltes}).
Each category describes the action the user performed in the SO answer edit, such as \texttt{fixing},  \texttt{updating}, and \texttt{improving}, and the target of the edit, such as \texttt{formatting}, \texttt{code}, and \texttt{bug}.
We adopted this classification and mapped the action performed to the edit target to derive the final category. 
Nonetheless, this categorisation is based solely on edits made to SO answers, including both text and code. Thus, many of the categories are not applicable to our study (e.g., fixing typos). Moreover, it does not consider the application of the code on software projects. As a result, we employed open coding to create additional categories based on observations of the changes required to adopt the latest SO answer into the software projects.

\begin{figure}[H]
    \includegraphics[width=1\linewidth]{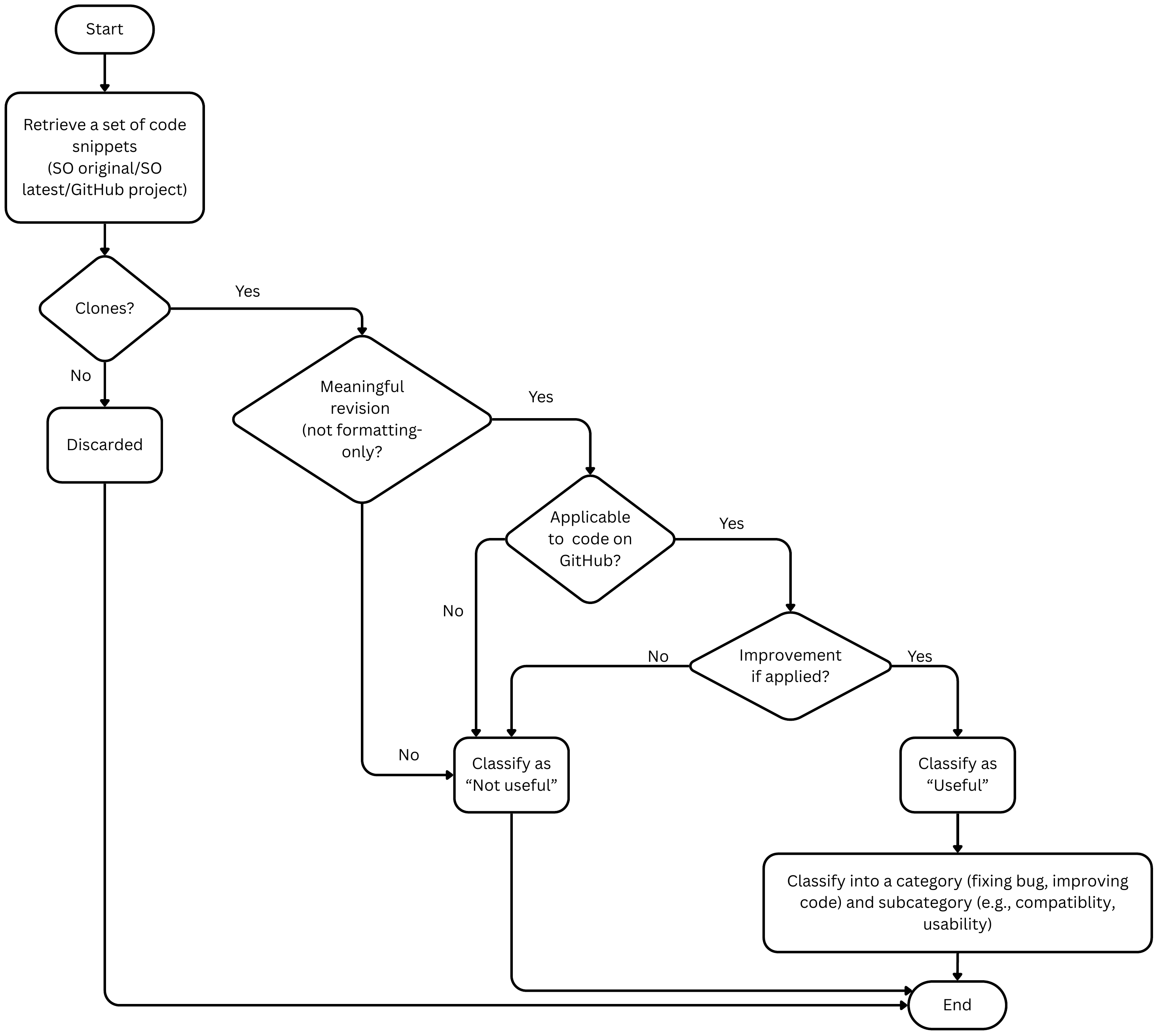}
\caption{{\bf Manual code book for classifying potential clones.}}
\label{fig:codebook}
\end{figure}

We followed a structured manual codebook to classify each snippet pair (as shown in Fig~\ref{fig:codebook}). First, we retrieved a set of code snippets containing the original version on SO, the latest version on SO suggested by Siamese+, and the associated GitHub project's code. We then verified the similarity between the original SO code and the latest SO code snippet (i.e., whether they were clones). If they are similar, we examined whether the latest SO revision introduced a meaningful change (not just formatting changes). Next, we checked whether it was applicable to the GitHub code. \rev{A recommendation was considered \emph{applicable} if the improvement introduced in the latest SO revision could also be applied to the corresponding GitHub code. For each matched GitHub--SO pair, we examined (i) whether the GitHub code and the SO code implemented the same functionality, (ii) whether they used the same or compatible APIs or libraries, and (iii) whether the proposed improvement could be applied without changing the intended behaviour of the original code. Only recommendations that satisfied all three criteria were marked as applicable and included in the subsequent analysis.} A pair was considered useful only if the revised snippet could be applied and resulted in a tangible improvement (e.g., bug fix or code quality enhancement). Otherwise, it was labelled as not useful.
Lastly, we classified them into a category following Baltes et al.~\cite{Baltes2020} and a subcategory.

\begin{table}[tb]
\caption{Categories of Stack Overflow post edits, following Baltes et al.~\cite{Baltes2020}.}
\label{tab:baltes}
\begin{tabular}{lp{8cm}}
\toprule
Type & Category \\
\midrule
Actions performed & adding, updating, deleting, fixing, improving, clarifying, simplifying, explaining, editing,
copy-editing, active reading, refactoring \\ 
\midrule
Target of the edit & formatting, typo, grammar, spelling, code,
bug, link, image, example, syntax, solution, tag \\
\midrule
Meta-level & sarcasm \\
\bottomrule
\end{tabular}
\end{table}

The manual classification was performed independently by two authors (first and fifth authors), who labelled each recommendation as applicable/non-applicable to their respective GitHub project and classified them into categories. 
After completing their independent classification, the authors compared their labelling, and the inter-rater reliability was measured using Cohen's kappa coefficient.
For recommendations with disagreements in classification, a third researcher \rev{with more than 20 years of Java programming experience} provided a final classification.
After the manual classification, we had a list of categories into which all potential clones had been categorised.
Finally, all SO's latest code answers that perform a relevant code update, such as \texttt{Fixing Bug} and \texttt{Improving Code}, were considered applicable to open-source projects. The manual validation supports this qualitative analysis.
The results obtained in this phase answer research questions RQ1 and RQ2\rev{, and they also yield the set of original--latest Stack Overflow answer pairs that we measure with static analysers in Phase~3 (Section~\ref{subsec:phase3}) to answer RQ3}.

\subsubsection{\rev{Phase 3: Measuring the answer edits with static analysis}}
\label{subsec:phase3}

\rev{%
The classification in Phase~2 rests on the judgement of human validators. To complement it with evidence that does not depend on human judgement, in this phase we measure the answer edits themselves, i.e., we quantify how the structure and the readability of a Stack Overflow code snippet change between its original and its latest revision. This phase answers RQ3.

\paragraph{Step 1 Scoping the snippet pairs:}
The input to this phase is the set of Stack Overflow answer revisions that Phase~2 matched to code in the GitHub projects. The 793 recommendations produced by Siamese+ do not correspond to 793 distinct answers, because the same answer can be recommended to several projects, and to several methods within the same project. We therefore deduplicated the recommendations by the answer's code block, which reduces them to 205 unique \emph{original--latest} pairs. Each pair consists of the original version of a code block, i.e., the root of its edit chain, and its most recent version, both of which were extracted in Section~\ref{subsec:sotorrent}. The comparison is thus paired by construction: every measurement is taken twice on the same code block, once before and once after the community's edits.

\paragraph{Step 2 Selecting and applying the static analysers:}
We used two tools, each targeting a distinct quality dimension, i.e., \textit{JavaParser}~\cite{javaparser,javaparser-visited} (version 3.28.1) for intrinsic structural measures and \textit{Checkstyle}~\cite{checkstyle} (version 13.9.0) for readability. We chose these two tools because they operate on the source code alone. Unlike tools that perform type resolution or inter-procedural analysis, neither requires the snippet to compile, to resolve its imports, or to have its dependencies available, none of which holds for code posted in a Stack Overflow answer.

From JavaParser we computed 19 structural measures, i.e., method, parameter, and local-variable counts; branch, loop, and switch counts; a documented cyclomatic-complexity proxy; maximum control-flow nesting; abrupt exits; try- and resource-handling counts; null-comparison counts; an aggregate exception-handling count; and the number of non-comment code lines. From Checkstyle we collected the findings of eight readability and style checks, which we report both as a raw finding count and as a density, i.e., findings per 100 non-comment lines, so that a snippet is not penalised merely for being longer.

Stack Overflow code snippets are frequently incomplete, e.g., a bare sequence of statements or a method without its enclosing class, which prevents a parser from processing them as they stand. To make such snippets analysable, we attempted each snippet under three wrapper strategies, i.e., the raw source as posted, the snippet wrapped as a class member, and the snippet wrapped as a method body. The strategies were applied \emph{symmetrically}, and a pair was analysed only when its original and its latest revision shared a common successful parse mode. This is a deliberately conservative rule: it discards pairs that could have been measured under mismatched wrappers, because a difference in the wrapper would otherwise be indistinguishable from a difference introduced by the edit itself. It is also the rule that determines how much of the sample each tool can cover, which we report in Section~\ref{sec:analysis-of-answer-edits}.

\paragraph{Step 3 Statistical analysis:}
Each measure was compared between the original and the latest revision with a two-sided \emph{paired Wilcoxon signed-rank test}. We selected this test because the two measurements of a pair are dependent, and because the distributions of the measures are heavily zero-inflated and skewed, i.e., most edits leave most counts untouched, which violates the assumptions of a paired $t$-test. To control the family-wise error rate across the many measures tested, we applied \emph{Holm's step-down correction}, over the 19 structural measures for JavaParser and over the two Checkstyle measures, and we report both the raw and the corrected $p$-values. Because a $p$-value alone says nothing about the size or the direction of a change, we also report the matched-pairs rank-biserial correlation $r_{\text{rb}}$ as an effect size, signed so that it expresses the change from the original to the latest revision, together with the number of pairs whose measure decreased, was unchanged, or increased. Finally, we repeated the structural analysis separately for the \texttt{Fixing Bug} and the \texttt{Improving Code} categories assigned in Phase~2, to test whether the two categories of edit differ in their measurable effect.

The results obtained in this phase answer RQ3 and are reported in Section~\ref{sec:analysis-of-answer-edits}.}

\subsubsection{\rev{Phase 4:} Submitting Stack Overflow latest code answers as pull requests}
\label{subsec:phase4}

\begin{figure}[H]
    \includegraphics[width=0.8\linewidth]{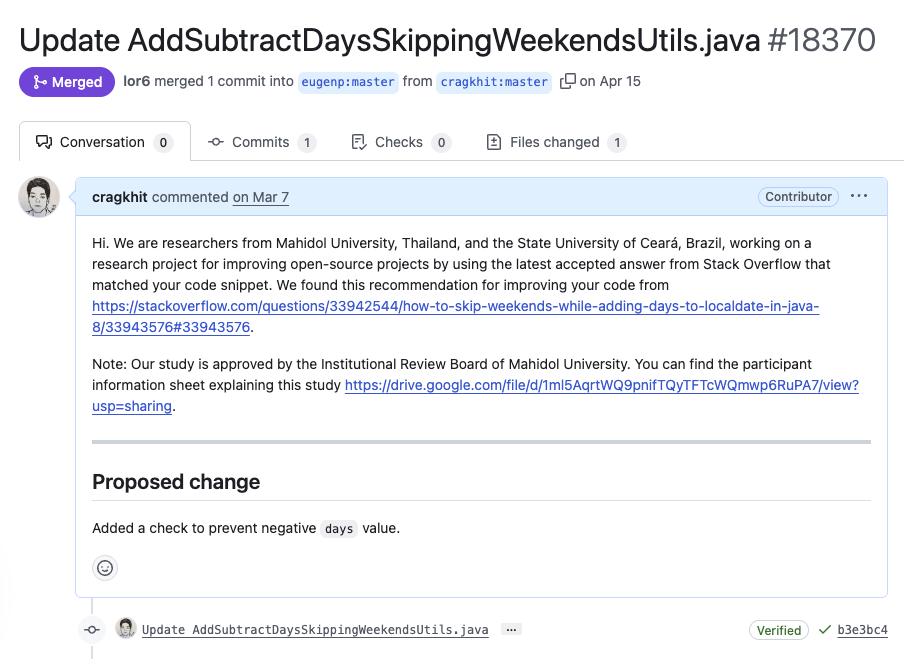}
\caption{{\bf Example of a pull request submitted to a GitHub project.}}
\label{fig:pr_example}
\end{figure}

To assess the usefulness of the SO's latest code answers identified in Phase 2 to open-source software developers, we created and submitted pull requests to their associated GitHub projects.
In the pull request, we made our best effort to implement minimal changes that incorporate the improvements based on SO's latest code answers.
The description for the pull request contains the explanation of this study and the ethics approval information. The information about the change was either adapted from the SO edit comment or created from scratch if no edit comment was provided. An example of such a pull request is shown in Fig~\ref{fig:pr_example}.

Then, we monitored each pull request's status and minimised interference with the review process whenever possible.
This is necessary to mitigate any bias we may introduce into the review process.

\paragraph{Ethics statement.}
\label{sec:ethics}
This study was reviewed and approved by the Mahidol University Central Institutional Review Board (MU-CIRB). 
The protocol number is MU-CIRB 2024/116.2203.
The study analysed publicly available Stack Overflow answers and public GitHub repositories, and it did not collect any personally identifiable information, so informed consent was not applicable.
Every pull request we submitted to a GitHub project disclosed the purpose of the study together with its ethics approval information, as described in Section~\ref{subsec:phase4}.

\section{Results}
\label{sec:results}
In this section, we discuss the findings from performing the \rev{four phases} of our experimental design and provide the answers to the \rev{three} research questions.

\subsection{\rev{Study setup}}
\rev{Before answering the research questions, we report two preparatory outcomes that configure our study pipeline: the optimised clone-search configuration obtained in Phase~1, and the filtered and grouped GitHub project dataset prepared in Phase~2.}

\subsubsection{The optimised Siamese configurations}
 
Given our defined grid, the Grid Search algorithm evaluated a total of \rev{3,072 configurations, plus Siamese's default configuration as a baseline}.
This process was completed in 54 hours, with each configuration taking an average of 63 seconds.
Tuning Siamese's parameters using the SO data (72,365 Java code snippets) and the Qualitas dataset of 111 Java projects resulted in the optimised parameters shown in Table~\ref{tab:optimized_config}, which offer the highest MRR score of 0.782.

Examination of the results for the configuration depicted in Table~\ref{tab:optimized_config} reveals several interesting facts.
First, Siamese never returned an empty set as results, i.e., for all queries, a set of candidate snippets was always returned.
Moreover, the correct clone was always present in the returned set of code snippets.
Although the correct snippet was not always displayed in the first position, the resulting MRR of 0.782 indicates that the correct clone was typically ranked at or near the top of the result list. This optimised parameter helps reduce the number of false clones detected by Siamese and the time required for the manual validation step. 
We utilise this tuned configuration in Siamese+ to identify similar SO answers and open-source Java projects on GitHub.


\begin{table}[tb]
\caption{Optimised configurations of Siamese.}
\begin{tabular}{ll}
\toprule
Parameters & Values \\
\midrule
Clone size & 6 lines \\
Size of n-gram & 1 (for representation r0) \\
 & 4 (for representation r1, r2, r3) \\
QR thresholds & 9, 6, 5, 9 (r0, r1, r2, r3) \\
Similarity thresholds & 50\%,60\%,70\%,80\% (r0, r1, r2, r3) \\
\bottomrule
\end{tabular}
\label{tab:optimized_config}
\end{table}




\subsubsection{Filtered and grouped GitHub projects}
After querying GHS (GitHub Search)~\cite{Dabic:msr2021data} using the following filters: \texttt{Language: Java; Exclude Forks; Has Open Issues; Has Open Pull Requests}, we obtained 20,976 GitHub projects.\footnote{\rev{This data collection was performed in March 2023.}} Then, we analysed the projects based on the three popularity metrics: \texttt{Number of Stars}, \texttt{Number of Watchers}, and \texttt{Number of Forks}. 

\begin{figure}[H]
    \begin{subfigure}[t]{0.45\textwidth}
        \centering
        \includegraphics[width=\textwidth]{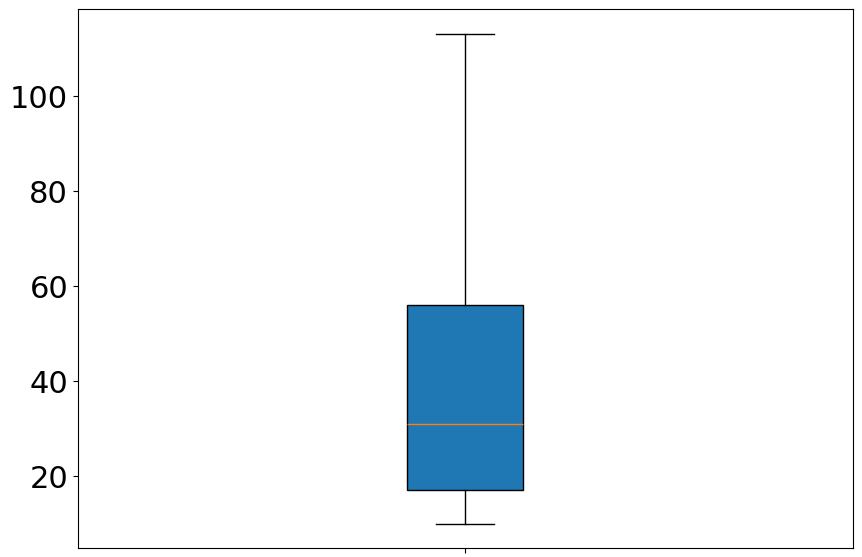}
        \caption{Stars}
    \end{subfigure}%
    ~
    \begin{subfigure}[t]{0.45\textwidth}
        \centering
        \includegraphics[width=\textwidth]{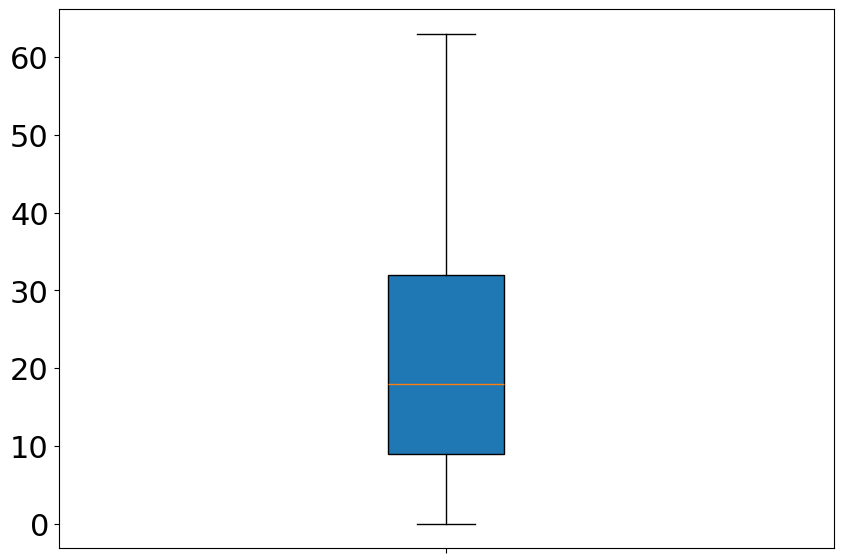}
        \caption{Forks}
    \end{subfigure}
    ~
    \begin{subfigure}[t]{0.45\textwidth}
        \centering
        \includegraphics[width=\textwidth]{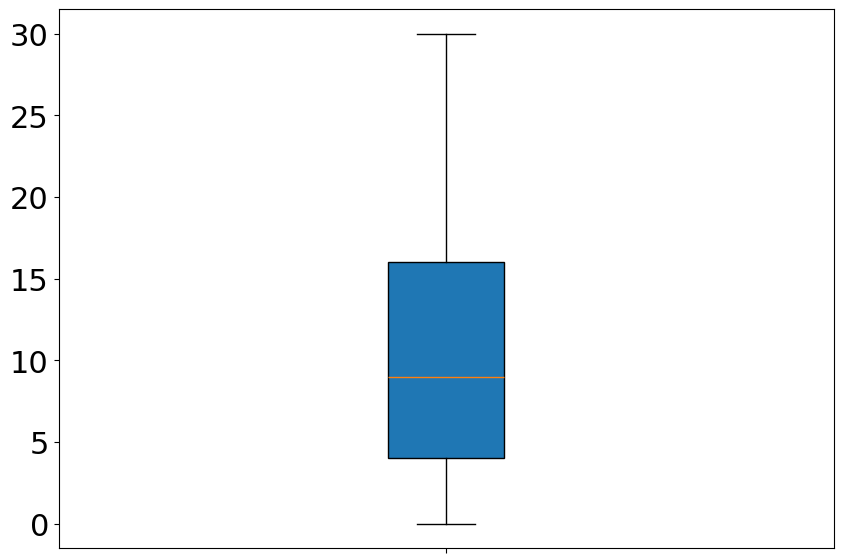}
        \caption{Watchers}
    \end{subfigure}
\caption{{\bf Distribution of the metrics of the collected GitHub projects.}}
\label{fig:gh_boxplot}
\end{figure}

The statistics of the collected GitHub projects are shown in Table~\ref{tab:summary_statistics} and displayed as boxplots in Fig~\ref{fig:gh_boxplot}. For stars, the values range from 10, the minimum value of stars of projects in the GHS system, to 145,397. The first, second, third, and fourth quartiles are between [10, 25], (25, 70], (70, 272], and (272, 145397] respectively. For forks, the values range from 0 to 50,497. The first, second, third, and fourth quartiles are between [0, 14], (14, 34], (34, 106], and (106, 50497] respectively. For watchers, the values range from 0 to 5,442. The first, second, third, and fourth quartiles are between [0, 6], (6, 13], (13, 31], and (31, 5442] respectively.

\begin{table}[tb]
    \caption{Summary statistics for stargazers, forks, and watchers.}
    \begin{tabular}{lrrr}
        \toprule
        & \textbf{Stars} & \textbf{Forks} & \textbf{Watchers} \\
        \midrule
        Mean & 558.80 & 185.72 & 38.93 \\
        Std. & 2,609.89 & 957.41 & 126.28 \\
        Min & 10.00 & 0.00 & 0.00 \\
        25\% & 25.00 & 14.00 & 6.00 \\
        50\% (Median) & 70.00 & 34.00 & 13.00 \\
        75\% & 272.00 & 106.00 & 31.00 \\
        Max & 145,397.00 & 50,497.00 & 5,442.00 \\
        \bottomrule
    \end{tabular}
    \label{tab:summary_statistics}
\end{table} 

\begin{figure}[H]
\centering
    \begin{subfigure}[t]{0.5\textwidth}
        \includegraphics[width=\textwidth]{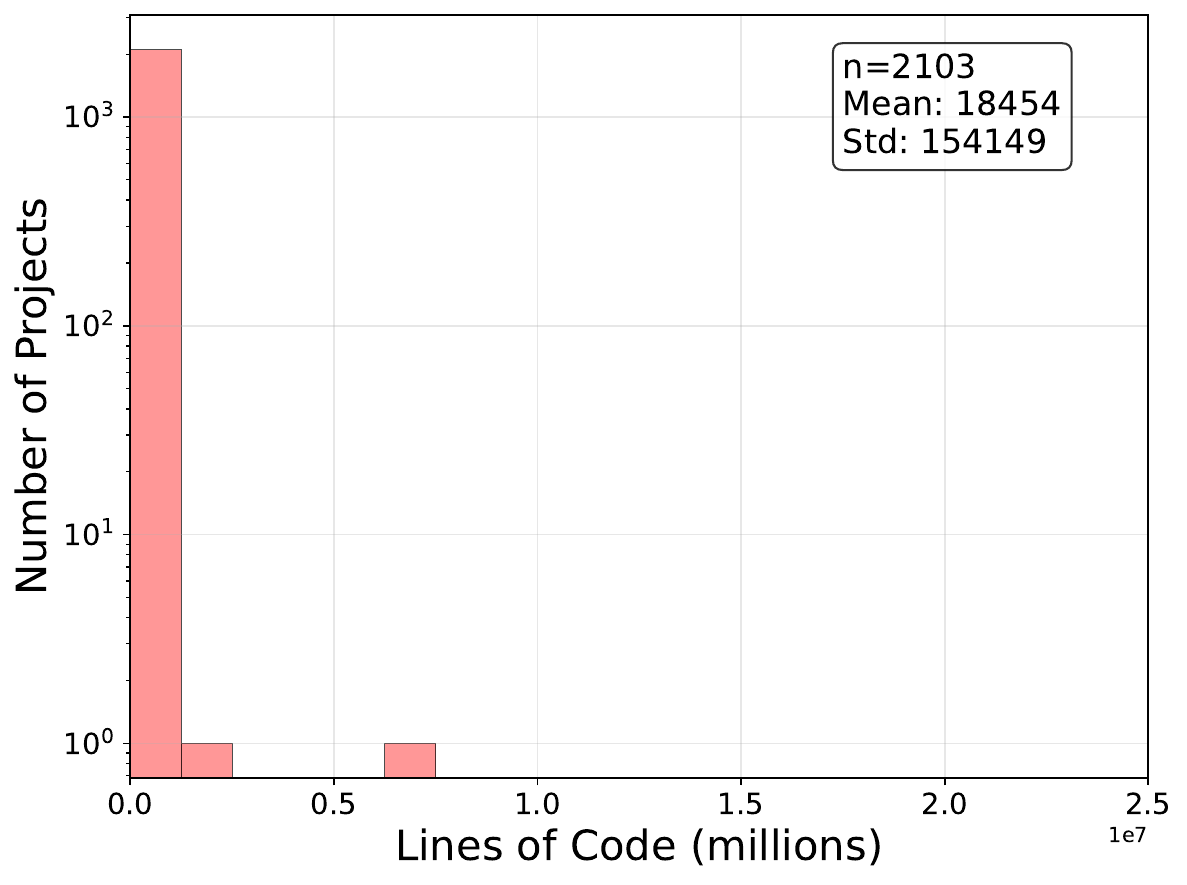}
        \caption{Low-popularity projects}
    \end{subfigure}%
    ~
    \begin{subfigure}[t]{0.5\textwidth}
        \includegraphics[width=\textwidth]{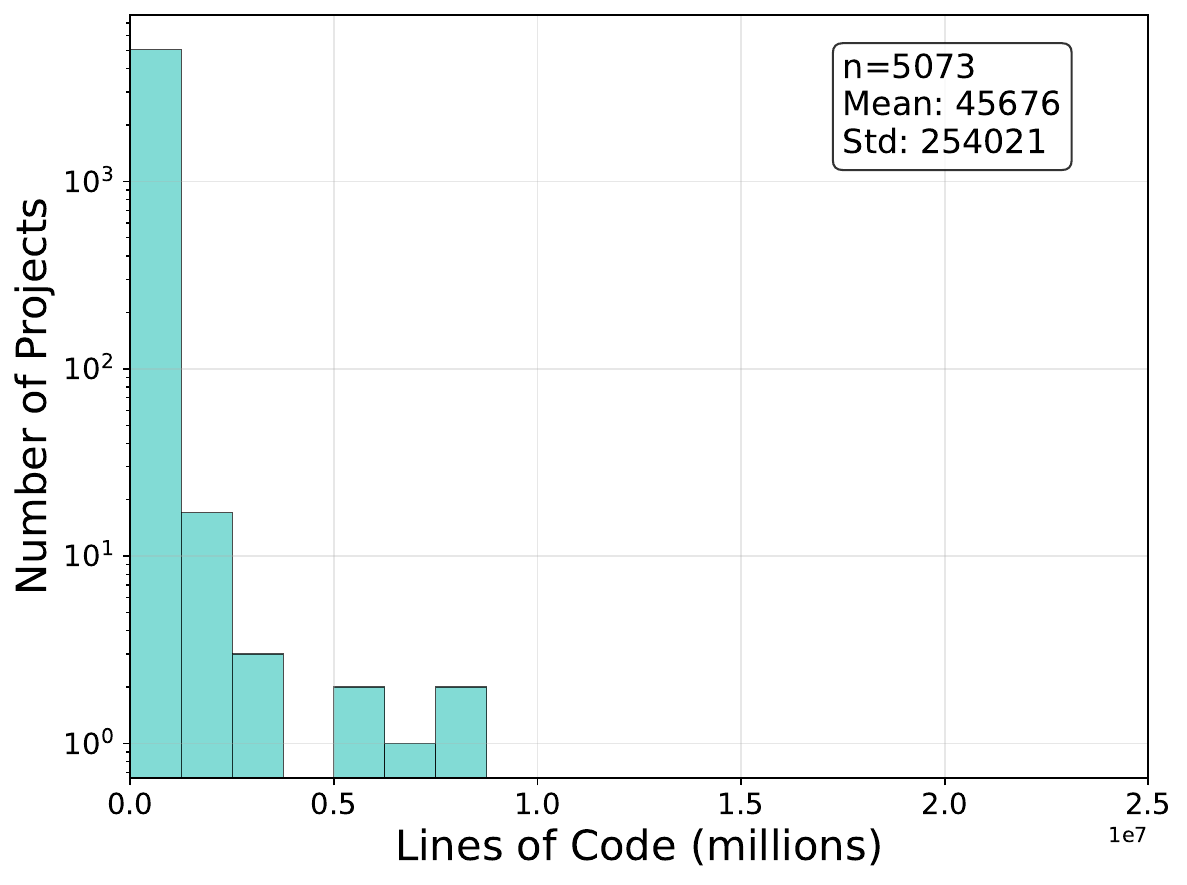}
        \caption{Medium-popularity projects}
    \end{subfigure}
    ~
    \begin{subfigure}[t]{0.5\textwidth}
        \includegraphics[width=\textwidth]{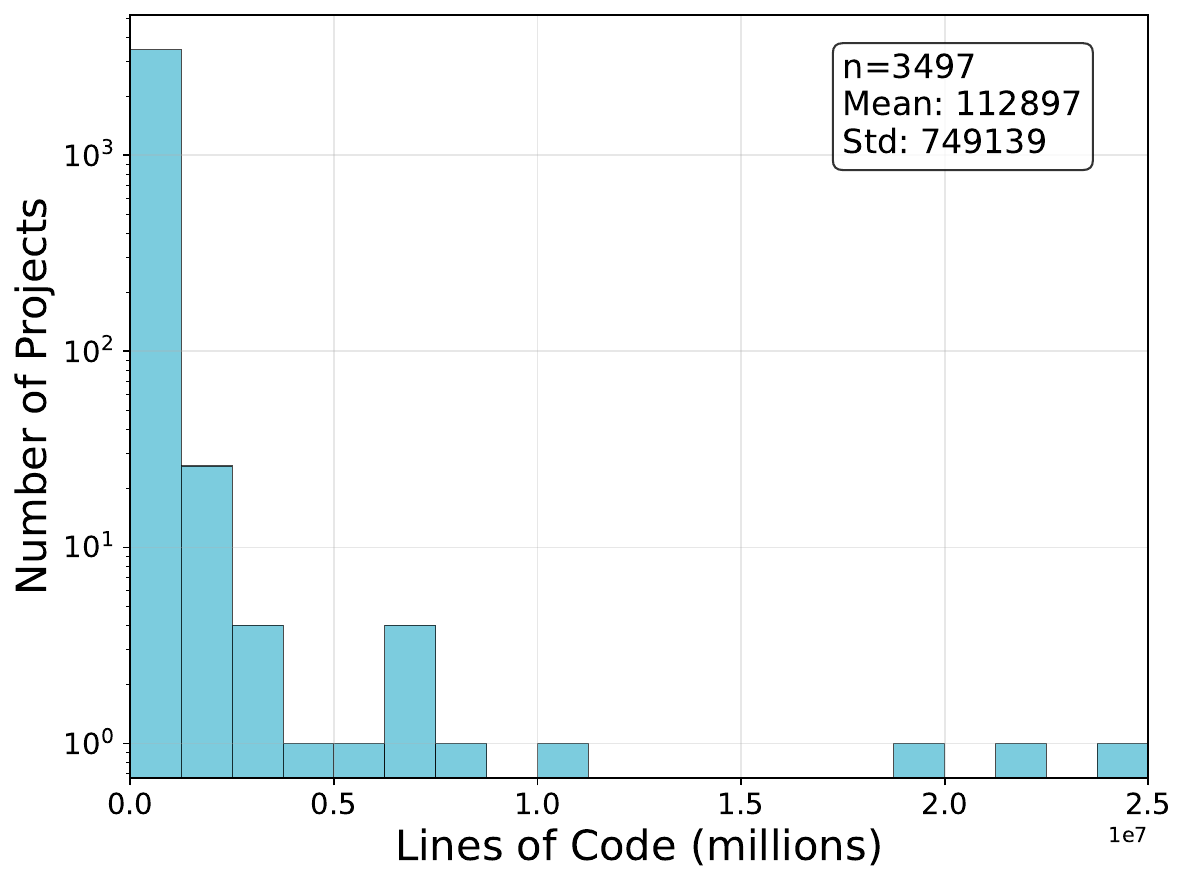}
        \caption{High-popularity projects}
    \end{subfigure}
\caption{{\bf Distribution of the code lines in each project group.}}
\label{fig:gh_size_boxplot}
\end{figure}

We filtered only the projects that passed the criteria articulated in Section \ref{sec:filter}, i.e., (1) the three metric values are all in the first quartile (low-popularity), (2) the three metric values are all in the fourth quartile (high-popularity), and (3) the three metric values are within the second and the third quartiles (medium-popularity).
This resulted in 10,673 GitHub projects containing 4,571,660 Java files. The number of projects in each group is shown in Table~\ref{tab:gh_groups}. The number of projects in the medium-popularity group is the highest (5,073), followed by the high-popularity (3,497) and low-popularity (2,103) groups. Moreover, the code sizes of the projects range from the low-popularity group (average code lines of 18k) to the medium-popularity group (46k) and the high-popularity group (113k), respectively. Fig~\ref{fig:gh_size_boxplot} shows the distribution of the project sizes in terms of the number of code lines in the three groups. As previously discussed, the number of code lines is largest in the high-popularity projects, indicating that they have larger code bases than the low- and medium-popularity groups. These GitHub projects were used to search for the latest code answers based on the SO data.

\begin{table}[tb]
\caption{Collected GitHub projects categorised by groups.}
\label{tab:gh_groups}
\begin{tabular}{lrrr}
\toprule
Group & Amount & Avg. Lines & Std. Dev. \\
\midrule
Low-popularity &  2,103 & 18,453.85 & 154,149.35 \\
Medium-popularity & 5,073 & 45,675.51 & 254,021.39 \\
High-popularity & 3,497 & 112,896.53 & 749,139.46 \\
\bottomrule
\end{tabular}
\end{table}





\subsection{RQ1: \rev{To what extent are accepted Java answers on Stack Overflow edited?}}
\label{subsec:rq1}

\begin{figure}[H]
    \includegraphics[width=0.9\linewidth]{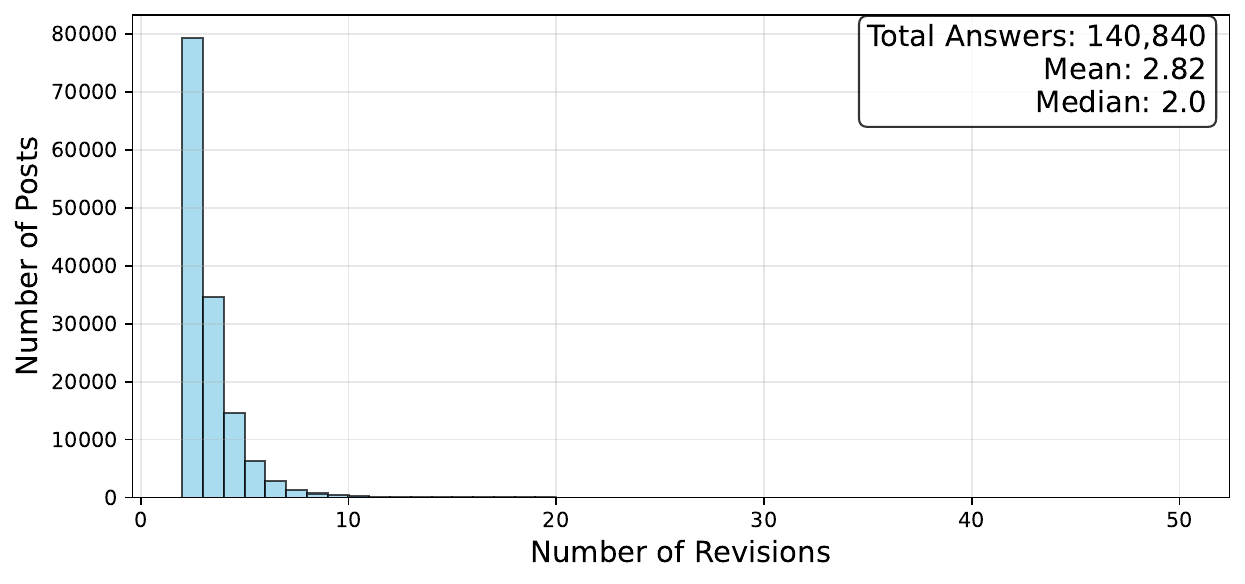}
\caption{{\bf The distribution of revisions in our filtered Stack Overflow answers.}}
\label{fig:so_edit_graph}
\end{figure}

\rev{We identified 2,832,800 Java answers on SO, sourced from SOTorrent, based on the \texttt{java} tag. From these answers, we filtered them to only accepted answers and obtained 874,438 answers which contains 917,389 Java code snippets. This resulted in an average of 1.05 code snippets per Java accepted answer, with a standard deviation of 1.22.}
From these accepted answers, we filtered only those that had at least one revision (i.e., had been edited at least once) and revisions to the code blocks. This resulted in 140,840 answers (\rev{16.11\% of the 874,438 accepted answers}) that had more than one revision. The average number of revisions per answer was 2.82, with a median of 2.0. The total number of code snippets across all revisions of accepted answers was 283,838. The distribution of the revisions in our SO Java accepted answers is shown in Fig~\ref{fig:so_edit_graph}. We indexed these 283,838 answer revisions in Siamese+. The answer with the most revisions among our collected SO answers had 53 revisions. 

The accepted answer with 53 revisions is the post ID 62805030\footnote{\url{https://stackoverflow.com/posts/62805030/revisions}}. It provides an explanation of multiple methods in the Apache Lang3 \texttt{StopWatch}, including \texttt{split()}, \texttt{stop()}, and  \texttt{getSplitTime()}. It also provides example code snippets for using such methods. The first revision was created on 8 July 2020 at 23:14. The answerer then continued updating their answer until the last revision on 24 January 2021 at 01:10 (spanning 6 months and 16 days).

\begin{figure}[H]
\centering
    \includegraphics[width=1\linewidth]{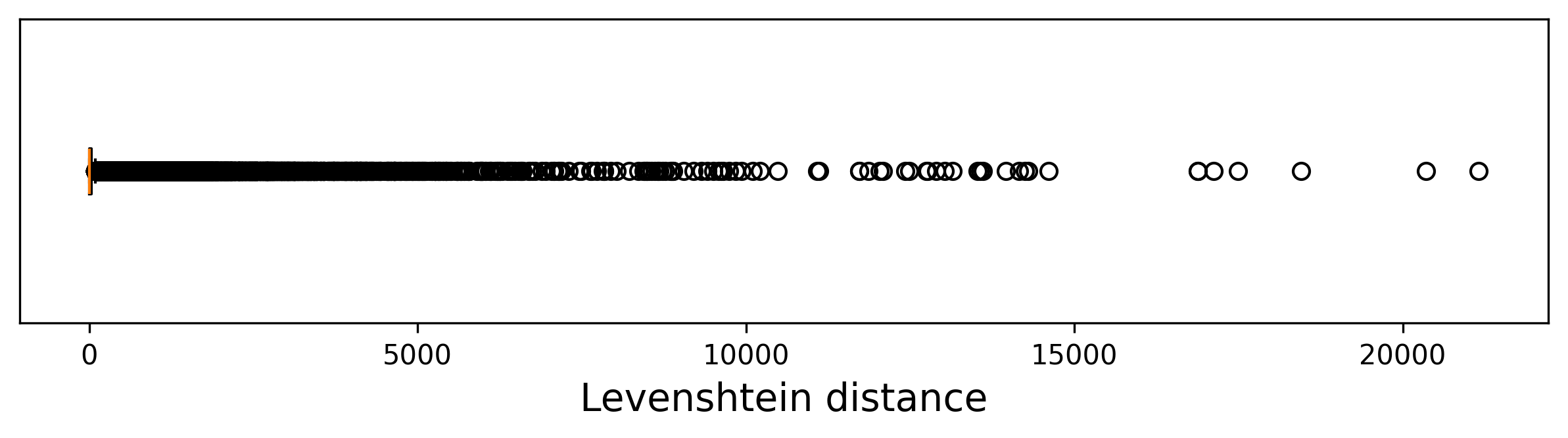}
\caption{{\bf Distribution of the Levenshtein distance between the original and the latest SO code answers.}}
\label{fig:distance}
\end{figure}

To gauge the size of code changes, we computed the Levenshtein distance, a metric that quantifies the similarity between two strings by counting the minimum number of single-character edits needed to transform one into the other. This comparison was made between the original code answer and their latest code answer to quantify the number of code changes. The distribution of the Levenshtein distance values is depicted in Fig~\ref{fig:distance} with the minimum distance of 0, maximum distance of 21,158 and the mean distance of 88.87 (standard deviation of 359.44). The accepted answer with the largest edit distance is post ID 24407816\footnote{\url{https://stackoverflow.com/posts/24407816/revisions}}, where the code of \texttt{JideScrollPane} in the answer was heavily modified over two revisions. So, most of the SO code answer edits are small, no more than 100 characters, but there are also some large edits that modify the majority of the code answer.

\begin{tcolorbox}
\textbf{Answer to RQ1:} \rev{16.11\%} of Stack Overflow Java accepted answers have more than one revision, with the average number of revisions per answer being 2.82. The average code edit size between the original and the latest SO revisions is 88.87 characters.
\end{tcolorbox}

\subsection{RQ2: \rev{What types of code improvements do Stack Overflow answer edits suggest for code found in open-source projects, and how are they distributed across projects of different popularity?}}
Siamese+ was used with the optimised configurations to generate recommendations for the 283,838 indexed SO answer revisions, using the 10,673 GitHub projects collected. The tool identified 793 pairs of code snippets in GitHub projects that included the latest code answers. Due to the large number of SO and GitHub code snippets, Siamese+ execution took 16 days, 5 hours, and 21 minutes.

\subsubsection{Classification of potential clones}
\label{subsubsec:classification}
\rev{The 793 recommended code snippet pairs were manually scrutinised by the second author, who had more than 8 years of code cloning research experience, to make sure that they were indeed code clones. The manual investigation revealed that the recommendations were all code clones. 
Nonetheless, the level of their code similarity varied, with some pairs being almost identical and others having significant differences. Based on the definitions of code clones by Roy et al.~\cite{Roy2009}, we found that 22\% of the pairs were Type-1 clones (i.e., identical code fragments except for variations in whitespace, layout and comments), 5\% were Type-2 clones (i.e., syntactically identical fragments except for variations in identifiers, literals, types, whitespace, layout and comments), while 73\% were Type-3 clones (i.e., code fragments with further modifications such as changed, added or removed statements). Thus, all recommended code snippet pairs were eligible for further analysis.}

\rev{Next,} the first and the fifth authors, who each had at least four years of Java programming experience, performed a manual investigation on the 793 pairs independently to (1) validate whether the recommendations are applicable to the open-source project or not (i.e., they actually improve the code in GitHub projects), and (2) classify them into one of the categories defined by~Baltes et al.~\cite{Baltes2020}. 
Code answers that performed trivial changes, such as adding comments, were classified as non-applicable. 
After the manual investigation, the two validators compared their results. There were 298 latest SO code answers that the two authors agreed to be applicable, 408 SO answers that the two validators agreed to be not applicable, and 87 SO answers with disagreements. This yielded a Cohen's Kappa score of 0.8333, indicating almost perfect agreement. Then, the two validators and \rev{the second author with more than 20 years of Java programming experience} discussed the applicability and classification of the 87 pairs for which there was disagreement until a consensus was reached. The final results contain 391 applicable recommendations and 402 non-applicable recommendations.

The classification results of the applicable latest SO code answers are shown in Table~\ref{tab:rec_types}. The 391 useful recommendations are mapped to either \textit{Fixing Bug} or \textit{Improving Code}. Our findings are similar to the study by Jallow et al.~\cite{Jallow2024}, which classified Stack Overflow answer edits into security-relevant, bug fixes, and improvements. Since the Improving Code category is coarse-grained, the two validators used open-coding to create an additional twelve subcategories of code improvement, including (1) application/framework specific, (2) code clarity and maintainability, (3) compatibility, (4) data and file handling, (5) data structures and algorithms, (6) error handling and robustness, (7) functional enhancement, (8) miscellaneous/others, (9) resource and process management, (10) security and cryptography, (11) usability, and (12) UI and interaction. The largest share of the Improving Code recommendations falls into \textit{code clarity and maintainability} (57), followed by \textit{application/framework-specific} (56), and \textit{data and file handling} (52).

\begin{table}[tb]
\caption{Results of manual type classification of the 391 applicable latest SO code answers.}
\begin{tabular}{llr}
\toprule
Baltes et al.~\cite{Baltes2020} category & Subcategory & Amount \\
\midrule
Fixing Bug & & 57 \\
\midrule
Improving Code & & 334 \\
\midrule
& Application/Framework-Specific	& 56 \\
& Code Clarity and Maintainability	& 57 \\
& Compatibility	& 7 \\
& Data and File Handling & 52 \\
& Data Structures and Algorithms & 14 \\
& Error Handling and Robustness & 45 \\
& Functional Enhancement & 21 \\
& Miscellaneous/Others & 21 \\
& Resource and Process Management	& 14 \\ 
& Security and Cryptography & 2 \\
& Usability	& 33 \\
& UI and Interaction & 12 \\
\bottomrule
\end{tabular}
\label{tab:rec_types}
\end{table}

\subsubsection{Distribution of the applicable SO's latest code answers}
\label{subsubsec:distribution}
\begin{figure}[H]
\centering
    \includegraphics[width=1\linewidth]{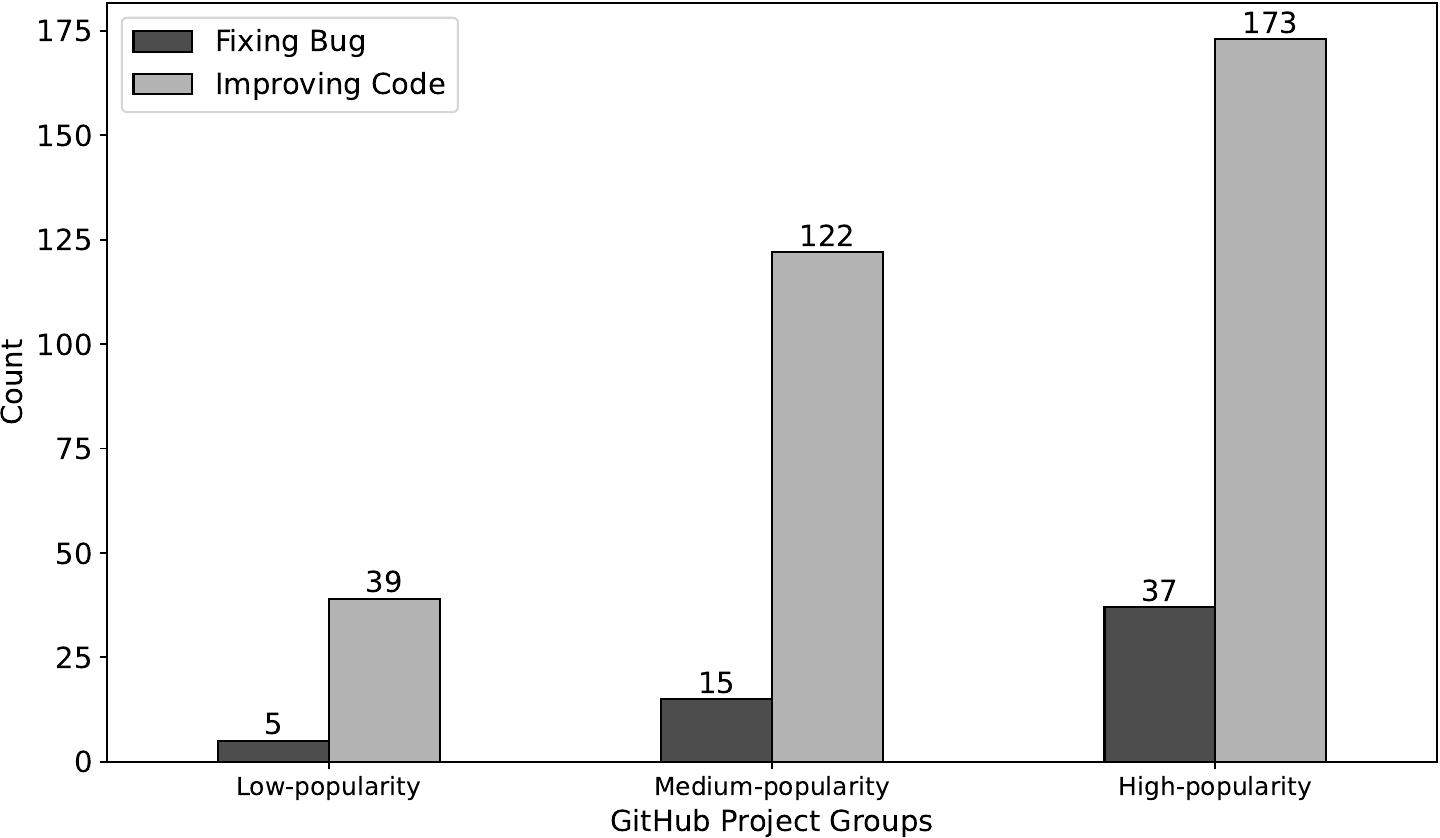}
\caption{{\bf Code recommendations grouped by GitHub projects.}}
\label{fig:recommendations_distribution}
\end{figure}

 As shown in Table~\ref{tab:gh_groups}, we analysed the distribution of the SO's latest code answers across the groups of the collected GitHub projects (low-popularity, medium-popularity, and high-popularity). We counted the occurrences of SO's latest code answers in the two categories (i.e., Fixing Bug and Improving Code) and divided them across the three GitHub groups. The results are depicted in Fig~\ref{fig:recommendations_distribution}. Examining the Fixing Bug category, we noted that the number of SO's latest code answers increased from the low-popularity group (5) to the medium-popularity group (15), and then to the high-popularity group (37). The same observation was found for the Improving Code category. As shown in Table~\ref{tab:recommendations_by_groups}, the maximum number of SO's latest code answers in the Fixing Bug category is around 1 to 12 (average values between 0.0024--0.0106). For the Improving Code category, the maximum number of SO's latest code answers is between 9 and 10 (average values between 0.0185 and 0.0495).

We performed a statistical test to assess differences in the number of SO's latest code answers in the Fixing Bug category across the three project groups. The null hypothesis ($H_0$) is \textit{There is no statistically significant difference in the number of Fixing Bug recommendations across the three groups}. The Shapiro-Wilk normality test~\cite{shapiro1965} suggested a non-normal distribution of the data across the three groups. Thus, we chose a non-parametric Kruskal-Wallis test~\cite{kruskal1952}. The test reported a p-value of 0.038673 ($\alpha$ = 0.05), a Kruskal-Wallis statistic (H) of 6.505224, and degrees of freedom of 2. Thus, we rejected the null hypothesis and concluded that there is a statistically significant difference between the Fixing Bug recommendations across the project quality groups.

We repeated the same test for the SO's latest code answers in the Improving Code category. The null hypothesis ($H_0$) is defined similarly as \textit{There is no statistically significant difference in the number of Improving Code recommendations across the three groups}. Since the data are not normally distributed, we again chose the non-parametric Kruskal-Wallis test. The test reported a p-value of 0.000016 ($\alpha$ = 0.05), a Kruskal-Wallis statistic (H) of 22.100237, and degrees of freedom of 2. We rejected the null hypothesis, concluding that there is also a statistically significant difference in the Improving Code recommendations across the three project quality groups.

The results indicate that the latest answers on SO, particularly those related to bug fixes and code improvements, are more applicable to high-popularity open-source projects than to medium- and low-popularity projects.

Since we identified 12 subcategories within the Improving Code category, we also analysed their distributions across the three GitHub project groups. The result is shown in Table~\ref{tab:improving_code_dist}, where the largest percentage value in each group is highlighted. For the low-popularity group, the largest SO latest answers fall into the application/framework specific subcategory (28.21\%). For the medium-popularity group, the largest portion (21.31\%) of the SO's latest answers is in the code clarity and maintainability subcategory. Lastly, for the high-popularity group, the largest portion (19.08\%) is for improving the code's error handling and robustness.

\begin{table}[tb]
\caption{Number of SO latest code answers grouped by GitHub project groups. Max and Avg. are the maximum and average number per project, respectively.}
\label{tab:recommendations_by_groups}
\begin{tabular}{lrrrrrr}
\toprule
Group & \multicolumn{3}{c}{Fixing Bug} & \multicolumn{3}{c}{Improving Code} \\
\cmidrule{2-4}\cmidrule{5-7}
& Total & Max & Avg. & Total & Max & Avg. \\
\midrule
Low-popularity & 5 & 1 & 0.0024 & 39 & 10 & 0.0185 \\
Medium-popularity & 15 & 1 & 0.0030 & 122 & 9 & 0.0240 \\
High-popularity & 37 & 12 & 0.0106 & 173 & 9 & 0.0495 \\
\bottomrule
\end{tabular}
\end{table}

\begin{table}[htbp]
\centering
\caption{Distribution of the Improving Code items across GitHub project groups.}
\label{tab:improving_code_dist}
\resizebox{\textwidth}{!}{%
\begin{tabular}{lrrr}
\toprule
Category & Low-popularity & Medium-popularity & High-popularity \\
& ($N=39$) & ($N=122$) & ($N=173$) \\
\midrule
Error Handling \& Robustness & 3 (7.69\%) & 9 (7.38\%) & \textbf{33 (19.08\%)} \\
Code Clarity \& Maintainability & 6 (15.38\%) & \textbf{26 (21.31\%)} & 25 (14.45\%) \\
Application/Framework-Specific & \textbf{11 (28.21\%)} & 21 (17.21\%) & 24 (13.87\%) \\
Data \& File Handling & 7 (17.95\%) & 25 (20.49\%) & 20 (11.56\%) \\
Usability & 3 (7.69\%) & 15 (12.30\%) & 15 (8.67\%) \\
Functional Enhancement & 1 (2.56\%) & 6 (4.92\%) & 14 (8.09\%) \\
Resource \& Process Management & 1 (2.56\%) & 1 (0.82\%) & 12 (6.94\%) \\
Data Structures \& Algorithms & 0 (0.00\%) & 4 (3.28\%) & 10 (5.78\%) \\
UI \& Interaction & 2 (5.13\%) & 1 (0.82\%) & 9 (5.20\%) \\
Miscellaneous / Others & 5 (12.82\%) & 8 (6.56\%) & 8 (4.62\%) \\
Compatibility & 0 (0.00\%) & 4 (3.28\%) & 3 (1.73\%) \\
Security \& Cryptography & 0 (0.00\%) & 2 (1.64\%) & 0 (0.00\%) \\
\bottomrule
\end{tabular}
}
\end{table}

\begin{tcolorbox}
\textbf{Answer to RQ2:}~49.31\% (391 out of the 793) of SO's latest code answers are considered applicable to the GitHub projects. The number of recommendations for both the Fixing Bug and Improving Code categories is the highest in high-popularity GitHub projects.
\end{tcolorbox}


\subsection{\rev{RQ3: What are the differences in terms of code structure and readability introduced by Stack Overflow answer edits?}}
\label{sec:analysis-of-answer-edits}

\rev{%
Following the procedure described in Phase~3 (Section~\ref{subsec:phase3}), we compared every original Stack Overflow snippet against its most recent revision, over the 205 unique original--latest pairs to which the 793 Siamese+ recommendations reduce, using \textit{JavaParser} for structural measures and \textit{Checkstyle} for readability.

We first report how much of that sample each tool could analyse. Table~\ref{tab:ba-coverage} shows the outcome of the symmetric wrapper procedure of Phase~3, under which a pair is measured only when its original and its latest revision share a common successful parse mode.
Both tools independently exclude roughly a third of the scoped sample because the original and the latest snippet never shared a common successful parse mode.

\begin{table}[htbp]
\centering
\caption{Parsing coverage of the 205 original--latest Stack Overflow code snippet pairs.}
\label{tab:ba-coverage}
\begin{tabular}{lrrrrr}
\toprule
Tool & Scoped pairs & Analysed & Excluded & Exclusion rate \\
\midrule
JavaParser & 205 & 128 & 77 & 37.6\% \\
Checkstyle & 205 & 129 & 76 & 37.1\% \\
\bottomrule
\end{tabular}
\end{table}

\subsubsection{JavaParser (structural metrics)}
Each of the 19 structural measures listed in Phase~3 was tested independently via a paired Wilcoxon signed-rank test, with Holm's step-down procedure applied across all 19 tests to control the family-wise error rate.
Table~\ref{tab:or-javaparser} reports the results.

\begin{table}[htbp]
\centering
\caption{Paired JavaParser structural-metric comparison ($n=128$).
\emph{Dec.}/\emph{Unc.}/\emph{Inc.} denote the number of pairs whose metric decreased, was unchanged, or increased from the original to the latest revision.
$p$ is the raw two-sided Wilcoxon signed-rank $p$-value; $r_{\text{rb}}$ is the matched-pairs rank-biserial correlation (recent $-$ original); $p_{\text{Holm}}$ is Holm-corrected across all 19 metrics.}
\label{tab:or-javaparser}
\small
\begin{tabular}{lrrrrrrr}
\toprule
Metric & $\Delta$ median & Dec. & Unc. & Inc. & $p$ & $r_{\text{rb}}$ & $p_{\text{Holm}}$ \\
\midrule
Code lines                   & $0$ & 34 & 31  & 63 & 0.004 & $+0.337$ & 0.070 \\
Method count                 & $0$ & 10 & 92  & 26 & 0.026 & $+0.423$ & 0.406 \\
Parameter count              & $0$ & 9  & 102 & 17 & 0.025 & $+0.499$ & 0.406 \\
Max.\ parameters             & $0$ & 6  & 115 & 7  & 0.497 & $+0.242$ & 1.000 \\
Local-variable count         & $0$ & 10 & 91  & 27 & 0.011 & $+0.478$ & 0.180 \\
\texttt{if} count            & $0$ & 11 & 103 & 14 & 0.508 & $+0.160$ & 1.000 \\
Loop count                   & $0$ & 1  & 121 & 6  & 0.078 & $+0.821$ & 0.781 \\
Switch count                 & $0$ & 0  & 128 & 0  & --    & --       & --    \\
Catch count                  & $0$ & 3  & 117 & 8  & 0.123 & $+0.545$ & 1.000 \\
Cyclomatic-complexity proxy  & $0$ & 13 & 88  & 27 & 0.028 & $+0.398$ & 0.406 \\
Max.\ control nesting        & $0$ & 10 & 106 & 12 & 0.262 & $+0.285$ & 1.000 \\
Abrupt-exit count            & $0$ & 12 & 90  & 26 & 0.049 & $+0.367$ & 0.628 \\
Try count                    & $0$ & 3  & 116 & 9  & 0.064 & $+0.615$ & 0.704 \\
Try-with-resources count     & $0$ & 0  & 126 & 2  & 0.500 & $+1.000$ & 1.000 \\
Throw count                  & $0$ & 1  & 120 & 7  & 0.195 & $+0.556$ & 1.000 \\
Empty-catch count            & $0$ & 1  & 124 & 3  & 0.625 & $+0.500$ & 1.000 \\
Empty-block count            & $0$ & 2  & 121 & 5  & 0.297 & $+0.500$ & 1.000 \\
Null-check count             & $0$ & 6  & 118 & 4  & 0.922 & $+0.055$ & 1.000 \\
Exception-handling count     & $0$ & 5  & 110 & 13 & 0.048 & $+0.538$ & 0.628 \\
\bottomrule
\end{tabular}
\end{table}

From the table, we can see that, across all metrics, the majority of pairs are unchanged (e.g.\ switch count (128/128), \texttt{if} count (103/128), method count (92/128)), indicating that most revisions do not alter the snippet's structural shape.
No metric is statistically significant after Holm correction.
Code-line count is the closest to significance: it increases in 63 of 128 pairs against 34 decreases, giving a raw $p=0.004$ that corrects to $p_{\text{Holm}}=0.070$.
Six further metrics, method count, parameter count, local-variable count, the cyclomatic-complexity proxy, abrupt-exit count, and exception-handling count, cross the \emph{uncorrected} $p<0.05$ threshold, but none survives correction (all $p_{\text{Holm}}\geq0.18$).

The most consistent observation is every rank-biserial correlation is non-negative.
Revisions skew towards \emph{increasing} every structural count and towards decreasing none.
This suggests that Stack Overflow revisions mainly add code, for instance validation or defensive checks.

Looking at the results based on the SO latest code answers' categories, we found that both the Fixing Bug and Improving Code categories show a similar pattern to the overall analysis, with most pairs unchanged and no metric significant after Holm correction.
\rev{The two subgroups do not cover the whole analysed sample, because a Phase~2 category was assigned only to the recommendations judged applicable; the remaining analysed pairs derive from recommendations classified as non-applicable and therefore carry no category.}
Within the Fixing Bug subgroup ($n=12$), code-line count shows a raw p-value of $p=0.039$ (7/12 pairs increased, $r_{\text{rb}}=+0.778$). Within Improving Code ($n=65$), we observed similar findings with a raw p-value of $p=0.099$ (33/65 increased, $r_{\text{rb}}=+0.272$).
Every other structural metric in both subgroups is dominated by unchanged pairs and shows no separation from the null.

\subsubsection{Readability (Checkstyle)}
\label{sec:or-checkstyle}

Eight readability and style checks were applied to the $n=129$ analysed pairs.
Table~\ref{tab:or-checkstyle} reports the paired comparison for raw finding counts and for findings per 100 non-comment lines. Neither measure reaches significance, and both lean mildly towards improvement. Based on the raw count, 25 pairs improve against 20 that worsen, giving a rank-biserial of $r_{\text{rb}}=-0.141$ (negative indicates improvement). Based on the density measure (i.e., findings per 100 non-comment lines), 39 pairs improve against 22 that worsen, giving $r_{\text{rb}}=-0.228$.

\begin{table}[htbp]
\centering
\caption{Paired Checkstyle finding comparison ($n=129$).
$r_{\text{rb}}$ is the matched-pairs rank-biserial correlation, signed recent $-$ original, so a \emph{negative} value indicates fewer findings after the revision (i.e.\ improvement).
$p_{\text{Holm}}$ is Holm-corrected across both rows.}
\label{tab:or-checkstyle}
\small
    \resizebox{1\textwidth}{!}{%
    \begin{tabular}{lrrrrrr}
    \toprule
    Metric & Improved & Unchanged & Worsened & $p$ & $r_{\text{rb}}$ & $p_{\text{Holm}}$ \\
    \midrule
    Raw findings         & 25 & 84 & 20 & 0.414 & $-0.141$ & 0.414 \\
    Density (Findings/100 lines) & 39 & 68 & 22 & 0.122 & $-0.228$ & 0.243 \\
    \bottomrule
    \end{tabular}
    }
\end{table}

Looking at the results based on the SO latest code answers' categories, we found that Fixing Bug ($n=12$) shows 2 improved and 2 worsened pairs ($p=0.875$, $r_{\text{rb}}=+0.200$) and Improving Code ($n=66$) shows 10 and 10 ($p=0.430$, $r_{\text{rb}}=-0.205$), exactly balanced in counts.
The effect sizes show that this count-level balance conceals a small difference in magnitude. In Improving Code, the worsening pairs are outranked by the improving ones (hence the negative, improvement-leaning $r_{\text{rb}}$), while in Fixing Bug the sign reverses.

\subsubsection{Discussion}
\label{sec:or-discussion}

No measure, structural or stylistic, separates the latest revisions from the original Stack Overflow code snippets, and in both analyses the majority of pairs are unchanged on every measure.
The revisions are therefore predominantly localised, altering a specific statement, API call, or condition rather than restructuring the snippet.
This is what one would expect from the manual classification in Table~\ref{tab:rec_types}: an edit that replaces a deprecated API call, corrects an off-by-one condition, or closes a resource changes the behaviour of the code without changing the counts that JavaParser measures or the layout that Checkstyle inspects.

The two analyses nonetheless lean in opposite directions, and the combination is informative.
JavaParser shows a non-negative effect size for every metric with at least one changed pair, i.e., revisions add methods, parameters, local variables, branches, and exception-handling constructs more often than they remove them, whereas Checkstyle leans mildly (and non-significantly) towards fewer findings.
Revisions thus tend to make snippets slightly longer and slightly more branch-heavy, most plausibly by introducing the validation, null-handling, and try/catch logic that the original answers omitted, which is in line with Error Handling and Robustness being one of the largest Improving Code subcategories (45 items, Table~\ref{tab:rec_types}).

The practical implication is that conventional static-analysis metrics are a poor proxy for the usefulness of a Stack Overflow revision.
A metric-driven filter would treat many of these edits as neutral, and would actively penalise the defensive and error-handling additions that our validators judged to be genuine improvements, because such additions increase complexity and code-size counts.
Usefulness here is semantic, i.e., it lies in whether the revised snippet is correct, current, and appropriate for the surrounding code, which is precisely the judgement our manual classification (Section~\ref{subsubsec:classification}) and the pull-request evaluation (Section~\ref{sec:practical}) were designed to capture.
This reinforces our positioning of the approach as a semi-automated recommender whose candidates are reviewed by a human maintainer, rather than as an automated quality-improvement pipeline driven by metric thresholds.

Two caveats constrain the interpretation of these results.
First, roughly 37\% of the scoped pairs could not be analysed by either tool because the original and the revised snippet never shared a common successful parse mode. The analysed subset is therefore biased towards more complete, self-contained snippets, and the excluded fragments may behave differently.
Second, the snippets are short and the per-category samples are small, particularly Fixing Bug, so the tests have limited power to detect the modest effects that edits of this granularity would be expected to produce.
Consequently, the absence of significance here should be read as evidence that these revisions do not systematically restructure or reformat code, not as evidence that they do not improve it.
We demonstrate the latter point in Section~\ref{sec:practical}, where we show that a number of the revisions are accepted by GitHub project maintainers, despite the absence of clear structural or stylistic improvement.
}

\begin{tcolorbox}
\rev{\textbf{Answer to RQ3:} The majority of the SO's latest code answers do not change the code structure or readability of the original code snippets. However, the revisions tend to add defensive and error-handling logic, which increases structural counts without a significant change in readability.}
\end{tcolorbox}

\subsection{\rev{Practical application of the approach (Phase 4)}}
\label{sec:practical}

To assess the usefulness of the SO's latest code answers, we \rev{selected 107 of them for submission as pull requests to their respective GitHub projects: the 57 items in the Fixing Bug category and 50 items in the Improving Code category}. We refer to these SO's latest code answers, submitted as pull requests, as ``code recommendations.'' \rev{Not all of them could be submitted. We had to drop 27 because the target project or file no longer existed, or because the project code had evolved and no longer matched the code the recommendation applied to, leaving 80 recommendations that were actually submitted, grouped into 57 pull requests. We therefore report three distinct quantities throughout this section, i.e., the recommendations \emph{selected} (107), those \emph{submitted} (80), and those \emph{accepted} (19), because the acceptance rate differs depending on which is used as the denominator. Table~\ref{tab:pr_funnel} summarises the funnel.}

\begin{table}[tb]
\caption{\rev{Phase 4 pull-request funnel. ``Selected'' counts the code recommendations chosen for submission; ``submitted'' excludes those whose target project or file no longer existed or whose surrounding code had evolved; ``accepted'' counts recommendations that were merged or that propagated into maintainer-applied changes. The acceptance rate is computed over the recommendations actually submitted.}}
\label{tab:pr_funnel}
\begin{tabular}{lrrr}
\toprule
\rev{Stage} & \rev{Fixing Bug} & \rev{Improving Code} & \rev{Total} \\
\midrule
\rev{Recommendations selected} & \rev{57} & \rev{50} & \rev{107} \\
\rev{Excluded before submission} & \rev{10} & \rev{17} & \rev{27} \\
\rev{Recommendations submitted} & \rev{47} & \rev{33} & \rev{80} \\
\rev{Pull requests opened} & \rev{24} & \rev{33} & \rev{57} \\
\rev{Pull requests accepted} & \rev{4} & \rev{8} & \rev{12} \\
\rev{Recommendations accepted} & \rev{11} & \rev{8} & \rev{19} \\
\midrule
\rev{Acceptance rate (of submitted)} & \rev{23.40\%} & \rev{24.24\%} & \rev{23.75\%} \\
\bottomrule
\end{tabular}
\end{table}

The \rev{57} Fixing Bug recommendations offer clear benefits in improving the code and reducing the burden on OSS project maintainers. However, we had to ignore ten Fixing Bug recommendations because the GitHub projects or files no longer existed\rev{, leaving 47 recommendations to submit}. Moreover, some of the pull requests contain multiple code recommendations for the same project, so we grouped them into a single pull request. Thus, finally, we opened 24 pull requests \rev{carrying these 47 recommendations}.
The details of the pull requests \rev{in this category} are shown in Table~\ref{tab:accepted_pr}. Looking at the GitHub project associated with the pull requests, there are 12 high-popularity-group projects (marked with H in the table), eight medium-popularity-group projects (marked with M in the table), and four low-popularity-group projects (marked with L in the table).

\rev{For the Improving Code category, we selected 50 recommendations. However, we had to ignore 17 of them because of non-existent files, for similar reasons to the Fixing Bug category, and because the GitHub files had evolved and no longer matched the revised code. Thus, finally, we submitted the remaining 33 recommendations as 33 pull requests, one per recommendation. The details of the pull requests in this category are shown in Table~\ref{tab:additional_pr}. Looking at the GitHub project associated with the pull requests, there are 18 high-popularity-group projects (marked with H in the table), 13 medium-popularity-group projects (marked with M in the table), and 2 low-popularity-group projects (marked with L in the table).}

For each pull request, we studied the differences between the original code and the SO's latest code answer. We then cloned the project, created a new branch for the proposed update, and applied the changes based on the SO's latest code answer, making adjustments to align with the existing code. Lastly, we created a pull request that included the study's purpose (along with the ethics approval) and a brief explanation of the changes.

Since the creation of the first pull request (27 January 2025) to the time of writing of this paper, there have been \rev{12} pull requests containing \rev{19} code recommendations that were accepted by the GitHub project maintainers. \rev{This corresponds to 23.75\% of the 80 recommendations we actually submitted (19/80). Measured against all 107 recommendations we selected, i.e., including the 27 that could never be submitted, the rate is 17.76\% (19/107). We use the 80 submitted recommendations as the denominator for the headline rate, since a recommendation that was never submitted could not have been accepted or rejected by a maintainer.} \rev{There were 8 pull requests, containing 8 code recommendations, that were directly merged by the developers (marked as M in the tables), and 4 pull requests, containing 11 code recommendations, for which the proposed changes triggered the updates and propagated the changes to other pull requests or issues (marked as C$_p$ in the tables), which were eventually merged.}

For the 20 unmerged \rev{Fixing Bug} pull requests (see Table~\ref{tab:unmerged_prs}), 17 of them are still pending for review, 1 pull request was closed with a suggestion to follow the project's procedure (which we followed but did not get any further response), and 1 pull request was closed without any comment from the maintainer. Lastly, 1 pull request is currently under review (i.e., we have received a response from the reviewer, but it has not yet been merged).
\rev{For the 25 unmerged Improving Code pull requests, 20 of them are still pending for review, 1 pull request was closed with a suggestion to follow the project's procedure (which we followed but did not get any further response), and 2 pull requests were closed without any comment from the maintainer. Lastly, 2 pull requests are currently under review.}

\begin{table}[tb]
\caption{List of the 24 opened pull requests \rev{from the Fixing Bug category}, their associated GitHub projects, and their statuses. The H, M, and L indicate high-popularity, medium-popularity, and Low-popularity project groups. For the statuses, \rev{U, M, C, C$_p$ denote Under Review, Merged, Closed (without merging), and Closed (with propagating changes), respectively}.}
\label{tab:accepted_pr}
\resizebox{1\textwidth}{!}{%
\begin{tabular}{@{}llllll@{}}
\toprule
No & GitHub Project & stars/forks/watchers & Group & Pull Request & Status \\
\midrule
1 & actiontech/dble & 1005/314/75 & H & actiontech/dble/pull/3898 & U \\
2 & aionnetwork/aion & 337/113/83 & H & aionnetwork/aion/pull/1177 & U \\
3 & leapframework/framework & 43/19/10 & M & leapframework/framework/pull/69 & U \\
4 & allwefantasy/serviceframework & 549/271/164 & H & allwefantasy/ServiceFramework/pull/96 & U \\
5 & threerings/getdown & 481/122/47 & H & threerings/getdown/pull/281 & U \\
6 & bcgit/bc-java & 1868/1026/126 & H & bcgit/bc-java/pull/2012 & \rev{C$_p$} \\
7 & chrislacy/tweetlanes & 779/296/77 & H & chrislacy/TweetLanes/pull/443 & U \\
8 & clouway/oauth2-server & 44/21/24 & M & clouway/oauth2-server/pull/100 & M \\
9 & cryptoguardoss/cryptoguard & 88/27/7 & M & CryptoGuardOSS/cryptoguard/pull/35 & U \\
10 & cyriux/mpcmaid & 24/6/4 & L & cyriux/mpcmaid/pull/16 & U \\
11 & eugenp/tutorials & 31934/50497/1543 & H &  eugenp/tutorials/pull/18370 & M \\
12 & giloutho/logfly5 & 18/6/5 & L & giloutho/Logfly5/pull/18 & U \\
13 & linkedin/rest.li & 2251/517/190 & H & linkedin/rest.li/pull/1078 & U \\
14 & meituan-dianping/zebra & 2536/696/167 & H &  Meituan-Dianping/Zebra/pull/114 & U \\
15 & mycatapache/mycat2 & 1298/438/91 & H & MyCATApache/Mycat2/pull/824 & U \\
16 & mycatapache/mycat-server & 9533/3889/676 & H & MyCATApache/Mycat-Server/pull/2958 & U \\
17 & PGMDev/PGM & 152/86/17 & M & PGMDev/PGM/pull/1486 & C \\
18 & SCADA-LTS/Scada-LTS & 481/225/67 & H & SCADA-LTS/Scada-LTS/pull/3094 & C \\
19 & sergueik/selenium\_java & 21/11/5 & L & sergueik/selenium\_java/pull/431 & U \\
20 & teamapps-org/teamapps & 18/9/5 & L & teamapps-org/teamapps/pull/130 & U \\
21 & ukwa/webarchive-discovery & 100/24/23 & M & ukwa/webarchive-discovery/pull/321 & M \\
22 & vipshop/pallas & 240/83/24 & M & vipshop/pallas/pull/275 & U \\
23 & xnnyygn/xraft & 184/90/7 & M & xnnyygn/xraft/pull/46 & U \\
24 & zitmen/thunderstorm & 54/30/16 & M & zitmen/thunderstorm/pull/76 & U \\
\bottomrule
\end{tabular}
}
\end{table} 

\begin{table}[tb]
\caption{List of the 33 new pull requests from the Improving Code category, their associated GitHub projects, and their statuses. The H, M, and L indicate high-popularity, medium-popularity, and low-popularity project groups. For the statuses, U, M, C, C$_p$ denote Under Review, Merged, Closed (without merging), and Closed (with propagating changes), respectively.}
\label{tab:additional_pr}
\resizebox{1\textwidth}{!}{%
\begin{tabular}{@{}llllll@{}}
\toprule
No & GitHub Project & stars/forks/watchers & Group & Pull Request & Status \\
\midrule
1 & 1024-lab/smart-admin & 1689/590/46 & H & 1024-lab/smart-admin/pull/133 & U \\
2 & AbFab3D/AbFab3D & 66/26/15 & M & AbFab3D/AbFab3D/pull/20 & U \\
3 & aionnetwork/aion & 337/113/83 & H & aionnetwork/aion/pull/1180 & U \\
4 & alchitry/Alchitry-Labs & 24/10/2 & L & alchitry/Alchitry-Labs/pull/38 & U \\
5 & alibaba/spring-cloud-alibaba & 24834/7611/979 & H & alibaba/spring-cloud-alibaba/pull/4376 & M \\
6 & ambiverse-nlu/ambiverse-nlu & 205/45/14 & M & ambiverse-nlu/ambiverse-nlu/pull/55 & U \\
7 & ant-media/Ant-Media-Server & 2951/496/95 & H & ant-media/Ant-Media-Server/pull/7990 & U \\
8 & Anuken/Arc & 225/80/15 & M & Anuken/Arc/pull/207 & C \\
9 & apache/hugegraph-toolchain & 51/60/13 & M & apache/hugegraph-toolchain/pull/750 & U \\
10 & artisynth/artisynth\_core & 36/20/9 & M & artisynth/artisynth\_core/pull/20 & C$_p$ \\
11 & atilika/kuromoji & 855/125/67 & H & atilika/kuromoji/pull/143 & U \\
12 & ballerina-platform/ballerina-lang & 3066/686/169 & H & ballerina-platform/ballerina-lang/pull/44681 & U \\
13 & bcgit/bc-java & 1868/1026/126 & H & bcgit/bc-java/pull/2386 & C \\
14 & carlspring/s3fs-nio & 44/22/8 & M & carlspring/s3fs-nio/pull/973 & M \\
15 & chatty/chatty & 755/157/39 & H & chatty/chatty/pull/562 & U \\
16 & DaxiaK/MyDiary & 1456/248/84 & H & DaxiaK/MyDiary/pull/77 & U \\
17 & dietzm/GCodeInfo & 36/15/11 & M & dietzm/GCodeInfo/pull/17 & U \\
18 & encryptedsystems/Clusion & 121/43/19 & M & encryptedsystems/Clusion/pull/28 & U \\
19 & evgenyzinoviev/gravitydefied & 73/22/8 & M & evgenyzinoviev/gravitydefied/pull/13 & U \\
20 & freedomotic/freedomotic & 385/489/50 & H & freedomotic/freedomotic/pull/535 & U \\
21 & giginet/CCSocialShare & 10/4/4 & L & giginet/CCSocialShare/pull/8 & M \\
22 & Kickflip/kickflip-android-sdk & 634/221/82 & H & Kickflip/kickflip-android-sdk/pull/72 & U \\
23 & KnowageLabs/Knowage-Server & 352/203/47 & H & KnowageLabs/Knowage-Server/pull/986 & M \\
24 & liquibase/liquibase & 3466/1579/137 & H & liquibase/liquibase/pull/7870 & C$_p$ \\
25 & MesquiteProject/MesquiteCore & 87/25/22 & M & MesquiteProject/MesquiteCore/pull/135 & U \\
26 & mucommander/mucommander & 776/171/49 & H & mucommander/mucommander/pull/1500 & C \\
27 & p2abcengine/p2abcengine & 72/26/18 & M & p2abcengine/p2abcengine/pull/12 & U \\
28 & joshiejack/Harvest-Festival & 65/35/28 & M & joshiejack/Harvest-Festival/pull/233 & U \\
29 & pylerSM/XInstaller & 136/66/22 & M & pylerSM/XInstaller/pull/58 & U \\
30 & Red5/red5-server & 3176/974/293 & H & Red5/red5-server/pull/448 & U \\
31 & TheAlgorithms/Java & 49966/16495/2235 & H & TheAlgorithms/Java/pull/7550 & M \\
32 & tony19/logback-android & 1070/156/48 & H & tony19/logback-android/pull/446 & C$_p$ \\
33 & yahoo/elide & 926/222/58 & H & yahoo/elide/pull/3420 & U \\
\bottomrule
\end{tabular}
}
\end{table}

\begin{table}[tb]
  \caption{Statuses of the unmerged pull requests.}
  \label{tab:unmerged_prs}
  \begin{tabular}{lrr}
    \toprule
    Status & \rev{Fixing Bug} & \rev{Improving Code} \\
    \midrule
    Pending for review & 17 & \rev{20} \\
    Closed & 2 & \rev{3} \\
    Under review & 1 & \rev{2} \\
    \midrule
    Total & 20 & \rev{25} \\
    \bottomrule
  \end{tabular}
\end{table}

\subsubsection{\rev{Examples of the merged and rejected pull requests}}
We discuss some of the merged pull requests based on the code recommendations from our study.

\paragraph{\rev{Bug fixing pull requests:}} The pull request to the bcgit project, a high-popularity-group project with 1,868 stars, 1,026 forks, and 126 watchers, contains 8 recommendations\footnote{\url{https://github.com/bcgit/bc-java/pull/2012}}. All the changes are made within the \texttt{isEqual} method by introducing an array object comparison and a null check before the comparison to avoid a \texttt{NullPointerException}, as shown in Fig~\ref{fig:pr1}. After we introduced the first recommendation to the file \path{core/src/test/java/org/bouncycastle/crypto/test/PKCS12Test.java} in the pull request, the owner of the bcgit project agreed with the changes and mentioned that they would switch to use another method called \texttt{areEqual} instead (``\textit{Thanks for the PR. I have instead removed this \texttt{isEqual} method and referenced \texttt{org.bouncycastle.util.Arrays\#areEqual} instead.}''). We added seven more similar recommendations as comments in the pull requests, and all were accepted.\footnote{The full discussion of this pull request is in \nameref{S1_File}.}  Since this project has 1,026 forks, the old buggy version of the code snippets that could cause a \texttt{NullPointerException} was adopted in these forked projects as well, and our recommendations fixed these issues.

\begin{figure}[H]
\begin{lstlisting}[language=diff]
@@ -26,6 +26,13 @@ private boolean isEqual(
    byte[]  a,
    byte[]  b)
    {
+       if(a == b){
+            return true;
+       }
+       if(null == a || null == b){
+           return false;
+       }
+
    if (a.length != b.length)
    {
        return false;
    }
    for (int i = 0; i != a.length; i++)
    {
        if (a[i] != b[i])
        {
            return false;
        }
    }
    return true;
}
    \end{lstlisting}
\caption{{\bf Example of a Fixing Bug recommendation in the pull request of the bcgit project.}}
\label{fig:pr1}
\end{figure}

Another example of a merged PR is the oauth2-server project, which is classified as a medium-popularity-group project with 44 stars, 21 forks, and 24 watchers\footnote{\url{https://github.com/clouway/oauth2-server/pull/100}}.
We proposed a change in the clouway/oauth2/app/AppServer.java file as shown in Fig~\ref{fig:pr2}. The recommendation adds a null check before stopping the server. The PR was merged by the project maintainer in just 2 minutes.

\begin{figure}[H]
\begin{lstlisting}[language=diff]
@@ -43,12 +43,14 @@ protected Injector getInjector() {
    Runtime.getRuntime().addShutdownHook(new Thread() {
      @Override
      public void run() {
-       try {
-         server.stop();
-       } catch (Exception e) {
-         e.printStackTrace();
+        if (server != null) {
+          try {
+            server.stop();
+          } catch (Exception e) {
+            e.printStackTrace();
+          }
        }
      }
    });
  }
}
}
    \end{lstlisting}
\caption{{\bf Example of a \rev{Fixing Bug} recommendation in the pull request of the oauth2-server project.}}
\label{fig:pr2}
\end{figure}

\paragraph{\rev{Improving Code pull request:}} \rev{The pull request to the spring-cloud-alibaba project, a high-popularity-group project with 24,834 stars, 7,611 forks, and 979 watchers,\footnote{\url{https://github.com/alibaba/spring-cloud-alibaba/pull/4376}} updates the \texttt{GatewayConfig.java} file by configuring the gateway example to cache successful CORS preflight responses for one hour. This is a subtle improvement but is useful in a corner-case situation. The code change is shown in Fig~\ref{fig:pr3}. The project maintainer considered it useful and accepted the pull request and merged it into the project.}

\begin{figure}[H]
\begin{lstlisting}[language=diff]
@@ -109,6 +109,7 @@ public CorsWebFilter corsFilter() {
     config.addAllowedHeader("*");
     config.addAllowedMethod("*");
     config.addAllowedOriginPattern("*");
+    config.setMaxAge(3600L);
     source.registerCorsConfiguration("/**", config);
 
     return new CorsWebFilter(source);
    \end{lstlisting}
\caption{\rev{{\bf Example of an Improving Code recommendation in the pull request of the spring-cloud-alibaba project.}}}
\label{fig:pr3}
\end{figure} 

\paragraph{\rev{Reasons for rejected pull requests:}} \rev{We also analysed the reasons for the rejected pull requests. Although almost all of the closed pull requests did not receive any comment, there are 2 pull requests that were closed with reasons provided by the project maintainers. The first one is the pull request to the Arc project, a medium-popularity-group project with 225 stars, 80 forks, and 15 watchers.\footnote{\url{https://github.com/Anuken/Arc/pull/207}} The project maintainer rejected the pull request with the following comment: \textit{``This value will never be printed, and using a BigInteger makes it impossible to avoid allocations if the StringBuilder is ever extracted.''} The second one is the pull request to the bcgit project, a high-popularity-group project with 1,868 stars, 1,026 forks, and 126 watchers.\footnote{\url{https://github.com/bcgit/bc-java/pull/2386}} The project maintainer rejected the pull request with the following comment: \textit{``Appreciate the effort to improve the library. It's an interesting idea, although in this case it's really the wrong spot - it's a test class and the method should never be called with anything that's null. ... For these reasons, this one I will not merge, however I would be interested to see what else you come up with.''}}

\rev{Both rejections share a common reason that the recommendation is defensible in isolation, but does not fit the context of the code it was transplanted into. In the Arc case, the maintainer raised two objections. First, the faulty input that the recommendation guards against is unreachable in practice (``this value will never be printed''), so the change protects against a condition the project does not encounter. Second, the recommended solution violates a non-functional constraint of the project: Arc is a performance-sensitive game framework in which allocations are deliberately avoided, and the \texttt{BigInteger} introduced by the SO answer would make the surrounding code impossible to keep allocation-free. In the bcgit case, the maintainer likewise objected to the location of the change, i.e., a test class whose method is never called with \texttt{null}. Thus, it is the case that will never occur in the project, and the recommended null check is unnecessary.} 

\rev{These two cases point to a limitation of reusing SO answer edits in existing projects. An SO answer is written for a general, self-contained problem, whereas the matching code in a project sits within a specific context, such as test code or a code path with constrained inputs, that can make the improvement unnecessary. Notably, neither maintainer disputed the technical validity of the underlying SO answer, and the bcgit maintainer explicitly invited further contributions. This suggests that the rejections reflect a lack of contextual fit rather than a rejection of crowd-sourced recommendations in general, and that recommending SO answer edits to projects requires reasoning about the surrounding code and the project's non-functional constraints, not only about the textual match between the code snippets.}

\paragraph{\rev{Discussion:}}~\rev{Overall, we opened 57 pull requests across the two categories, of which 12 pull requests containing 19 code recommendations were accepted by the project maintainers, either by being merged directly or by triggering equivalent changes in the project. These pull request results demonstrate that the Stack Overflow code recommendations are perceived as useful by the open-source software developers and are accepted for merging into their software projects, even though they were submitted as unsolicited contributions from external parties. A few closed pull requests demonstrate that maintainers rarely dismiss the recommendations outright, and that rejection occurs when a recommendation does not fit the context of the receiving code rather than because the underlying Stack Overflow answer is incorrect.}\footnote{A few additional examples of the pull requests can be found in \nameref{S1_File}.} 

\subsubsection{Comparison to existing studies}
\rev{To put our acceptance rate into perspective, we compare it with related studies that also validate automated or semi-automated code improvements through real pull requests. In total, we opened 57 pull requests, i.e., 24 carrying Fixing Bug recommendations and 33 carrying Improving Code recommendations, of which 12 were accepted by the project maintainers. This corresponds to an acceptance rate of 21.1\% (12/57) at the level of pull requests, i.e., 16.7\% (4/24) for the Fixing Bug category and a higher 24.2\% (8/33) for the Improving Code category, and of 23.75\% (19/80) at the level of the individual code recommendations actually submitted. The most directly comparable study is that of Tang and Nadi~\cite{Tang2021}, who similarly mined Stack Overflow answer edits (from comment-edit pairs) and validated them by submitting 15 pull requests, of which 10 were accepted (66.7\%). Although their acceptance ratio is higher, their submissions were a small, hand-selected set targeting a few well-known repositories, whereas our pull requests span both bug fixes and general code improvements and were submitted to projects spanning different popularity levels.

Beyond Stack Overflow-based approaches, techniques that deliver automated or AI-assisted changes report substantially higher acceptance rates, but under conditions that strongly favour acceptance. For instance, the B-Assist automated program repair tool deployed at Bloomberg reached a 74.56\% acceptance rate by presenting patch suggestions in context within active pull requests to the engineers most familiar with the affected code~\cite{Williams2023}, and a recent study of agentic (LLM-generated) pull requests reports that 83.8\% were eventually merged~\cite{Watanabe2025}. 

These settings differ fundamentally from ours, in which improvements are proposed by an external party with no prior relationship to the project. This distinction matters because pull requests from external contributors are known to be rejected far more often than those from project insiders. Pinto et al.~\cite{Pinto2018} found that volunteers face roughly 26 times more rejection than employees contributing to the same projects. Our figures should also be read as a lower bound rather than a final outcome, since 37 of the 57 pull requests are still pending review and only five were closed without being adopted (see Table~\ref{tab:unmerged_prs}). The acceptance rate can therefore only increase as the remaining pull requests are reviewed. Viewed against these existing studies, and given that our submissions were unsolicited, externally submitted, and derived from Stack Overflow edits, our acceptance rates represent an encouraging result rather than an underperformance.}


\section{Discussion}

\rev{\subsection{Generalisability beyond Java: a replication on Python and JavaScript}
\label{subsec:replication}
To establish whether our findings are specific to the Java community or hold for Stack Overflow more broadly, we perform a smaller-scaled replication study on Python and JavaScript, the two most widely used languages alongside Java, in a companion technical report~\cite{Ragkhitwetsagul2026}. The replication deliberately changes as little as possible: it uses the same SOTorrent release (version 2020-12-31), the same data extraction pipeline, the same Siamese+ tool with the same tuned configuration, and the same popularity thresholds, so that any difference in the results can be attributed to the language rather than to the method. It analyses 840,132 accepted Python answers and 1,144,185 accepted JavaScript answers, and searches 100 GitHub projects per language, balanced across the three popularity groups.

Both findings generalise. For RQ1, the proportion of accepted answers carrying at least one edit is 41.25\% for Python and 39.10\% for JavaScript, against the 16.11\% we report for Java in Section~\ref{subsec:rq1}. The supply of potential code improvements is therefore roughly two and a half times larger, in proportional terms, in both replication languages than in Java, so the motivation for our approach is considerably stronger outside Java rather than weaker. 
The mean number of revisions per edited answer is 2.78 for Python and 2.68 for JavaScript against 2.82 for Java, and the median is exactly 2.0 in all three languages. 
The three communities differ substantially in \textit{how many} answers get edited, but barely at all in \textit{how much} an answer is edited once someone starts. This near-invariance is consistent with the editing behaviour being driven by a mechanism common to the platform, namely the suggested-edit and review workflow, whose typical outcome is one or two accepted revisions, rather than by anything specific to a language community.

For RQ2, the popularity effect replicates in direction in both languages. The number of matched answer edits rises monotonically from the low- to the medium- to the high-popularity group, from 80 to 156 to 977 for Python and from 32 to 71 to 353 for JavaScript, which is the same direction we report in Section~\ref{subsubsec:distribution} for both the Fixing Bug and the Improving Code categories. The Kruskal--Wallis test confirms the difference for Python ($p = 0.0018$), as it does for Java, but not for JavaScript ($p = 0.2175$). Nonetheless, we do not read the JavaScript outcome as a failure to replicate. The replication searched 100 projects per language against the 10,673 analysed here, leaving only 33 or 34 projects per group with a median of zero matched edits, so the test has very little power to detect anything short of an overwhelming effect. The popularity effect is thus directionally confirmed in both languages and statistically confirmed in one.


There are two limitations of this replication study. First, the replication performs no manual validation, so its counts are candidate matches rather than the manually confirmed applicable improvements we report for Java, and it consequently says nothing about what kind of improvement the matched edits represent. Second, it covers RQ1 and RQ2 only; RQ3 and the pull-request evaluation of Section~\ref{sec:practical} remain Java-only. We discuss the residual threat this leaves in Section~\ref{sec:threats}.}

\subsection{Implications for software engineering researchers}
The findings from this study corroborate those of related studies on answer edits on Stack Overflow~\cite{Tang2021,Zhang2022,Jallow2024}. We similarly found that outdated answers on SO are still being used in open-source software projects, even though newer versions address the issues of concern. 
\rev{The results from the practical application of our approach (Section~\ref{sec:practical}) indicate that at least some of} the identified latest SO code answers are applicable to real-world open-source software projects, \rev{with a subset of them accepted by the project maintainers}.
Future research can focus on developing an automated tool, based on the design of our Siamese+ tool, that recommends code improvements for open-source software projects using crowd-sourced SO answers. \rev{As noted in Section~\ref{subsec:replication}, such a tool should establish early that the code snippet itself, and not merely the surrounding prose, was changed by the edit, since roughly two thirds of the code snippets in edited answers are left untouched by them~\cite{Ragkhitwetsagul2026}.} Moreover, our manual classifications show that not all of the latest SO code answers are applicable to GitHub software projects. Thus, future studies can develop machine learning models or leverage LLMs to classify whether the latest SO code answers are applicable to software projects, thereby reducing the number of false-positive recommendations proposed to developers. 

\subsection{Implications for open-source software maintainers}
Our study and related studies reveal that code answers on SO change over time~\cite{Diamantopoulos2019,Tang2021,Sheikhaei2023, Mondal2025,Zhang2022,Jallow2024}. Some of the original answers may contain bugs or security weaknesses. Thus, it is important that open-source project maintainers keep track of code reused from SO, potentially by adding a comment with attribution and a link to the SO post or answer the code is copied from. By adding comments, they can easily track whether reused code from SO remains up to date. Moreover, they can adopt the Siamese+ tool used in this study to analyse their codebase against the SO answer revision data.

\subsection{Code recommendations from SO in the era of Generative AI}
With the rise of Generative AI (GenAI) tools, especially models specifically designed for code like GitHub's Copilot or Claude Code, generating code and fixing bugs have become \rev{considerably easier}. However, recent studies reveal that code snippets generated by these GenAI tools are not always high-quality~\cite{Liu2024,Wang2024,Fu2025,Siddiq2024,Shihan2024,Tambon2025,Cotroneo2025}. They may contain subtle bugs, maintenance issues, suboptimal code, or security vulnerabilities. Thus, it is important that the developers verify their correctness before integrating them into the code bases. We believe the findings from our study can complement this ongoing trend towards AI-assisted programming. By leveraging crowd-sourced knowledge from SO, developers can manually review the code generated by GenAI tools and verify it against the latest answers on the platform.


Notably, a recent study by Helic et al.~\cite{Helic2025} reports that the questions and answers on SO have evolved with the emergence of GenAI tools. The recent questions and answers on SO address more complex and challenging programming problems than in the past because they are the problems that cannot be handled by GenAI. Thus, we foresee that these code improvement recommendations based on SO will remain useful in the near future, particularly for complex code that cannot be fixed by GenAI.

\rev{\subsection{Limitations of leveraging Stack Overflow answer revisions}
While our results show that SO answer revisions can yield useful code improvements, leveraging them for software maintenance has inherent limitations along three dimensions. 

First, regarding code security, SO answer revisions are crowd-sourced and are not systematically vetted for security. An edit that fixes a functional bug does not necessarily address security weaknesses, and prior work has shown that widely reused SO snippets can contain vulnerabilities that persist across revisions~\cite{Verdi2020,Jallow2024}. Adopting the latest SO revision is therefore not a guarantee of a secure fix, and in the worst case an edit could introduce a new weakness. For this reason, our approach treats a revision as a recommendation to be reviewed by a human maintainer rather than as a patch to be applied automatically. 

Second, regarding project-specific requirements, SO answers are generic, context-free solutions to isolated programming problems, whereas real projects impose constraints such as specific API and library versions, coding conventions, performance and concurrency assumptions, and architectural decisions. An edit that is a genuine improvement in isolation may not fit a particular project, may conflict with its style, or may require non-trivial adaptation before it can be merged; this partly explains why not all applicable recommendations were accepted. 

Third, regarding evolving programming practices, both SO content and the surrounding language and library ecosystems evolve over time, so a ``latest'' SO revision reflects the best practice only at the time of that edit and may itself become outdated as new language features are introduced, APIs are deprecated, and conventions change. This limitation is compounded by the temporal coverage of our SOTorrent dataset (Section~\ref{sec:threats}). 

Taken together, these limitations define the setting in which our approach is most appropriate: a semi-automated recommender that surfaces candidate improvements for human review, rather than an autonomous maintenance tool.}

\section{Threats to validity}
\label{sec:threats}
There are potential threats to the validity of our study, which we categorise into \rev{construct, internal, conclusion, and external} threats as follows.

\rev{\textbf{Construct validity:}
A key assumption underlying our study is that a matched GitHub snippet is an adoption of an earlier Stack Overflow answer that has since been improved on SO. Our clone-based approach, however, establishes textual similarity between the two, not the direction or the origin of the reuse. A matched pair could instead reflect independent implementations of a common programming task, code copied from GitHub to SO, or reuse from a shared third-party source. We mitigated this threat by manually inspecting each candidate pair and retaining only those in which the latest SO revision constitutes a plausible, applicable improvement over the code in the GitHub project. Nevertheless, we cannot confirm the reuse direction for every pair, and our findings should be interpreted as evidence of applicable improvements rather than of confirmed SO-to-GitHub reuse. 

Moreover, we show the usefulness of a recommendation through the acceptance of its corresponding pull request. Pull-request acceptance is influenced by factors beyond code quality, such as the maintainers' availability and responsiveness, the project's activity level, its contribution guidelines, and the perceived reputation of the contributor. Consequently, a rejected or unanswered pull request does not necessarily indicate that a recommendation is not useful, and vice versa. Lastly, we use the Levenshtein distance between the original and latest answers as a proxy for the magnitude of a code change. As a character-level metric, it is sensitive to superficial edits such as formatting, whitespace, and identifier renaming, and may therefore not fully reflect the semantic significance of a change.}

\textbf{Internal validity:}
A potential threat to internal validity arises from the adoption of the Siamese code clone search tool to identify similar code snippets in SO accepted answers, which may yield false positives and false negatives. 
We mitigated this threat in Phase 1 of our study by searching for optimised configurations of the tool to reduce the number of false results.
Nonetheless, the optimised Siamese configuration is based on the Qualitas corpus, which dates back to 2013. Thus, it is possible that the optimised configurations may miss some of the Java language features and constructs released after 2013.
Additionally, GitHub's popularity metrics (i.e., stars, watchers, forks) may not accurately reflect a project's code quality. 
Nevertheless, we argue that a categorisation criterion combining all three metrics to create three project groups may be a more trustworthy proxy than relying on any single metric alone.
Another threat is that the manual validation step may contain errors due to human mistakes and biases. We mitigated this threat by employing two validators and cross-checking the results. \rev{The second author} was involved in resolving disagreements.
Moreover, the SO data used in this study is based on the SOTorrent dataset version 2020-12-31, which may differ from the latest Stack Overflow data.
\rev{Finally, because we both created the pull requests and interpreted their outcomes, the framing of our submissions, each of which disclosed the study's purpose and ethics approval, may have influenced the maintainers' willingness to review or merge them. We attempted to minimise this effect by keeping the pull-request descriptions factual and by not intervening in the review process, but a potential bias may remain.}

\rev{\textbf{Conclusion validity:}
We assessed the differences in the number of Fixing Bug and Improving Code recommendations across the low-, medium-, and high-popularity project groups using the Kruskal--Wallis test (Section~\ref{subsubsec:distribution}), after checking normality with the Shapiro--Wilk test. However, the finer-grained distribution of the twelve Improving Code subcategories across the groups is reported descriptively, and the pull-request acceptance outcomes are based on a small sample (57 pull requests, of which 12 have been accepted and 37 are still awaiting review) that we did not subject to significance testing. These latter results should therefore be interpreted as preliminary rather than as statistically confirmed effects.}

\textbf{External validity:}
The conclusions drawn from this study are based on the selection of GitHub projects according to our predefined methodology, detailed in Phase 2. 
Due to possible selection bias, the conclusions may not generalise to all GitHub projects. 
Moreover, we analysed only code written in Java across GitHub and SO answers.
\rev{We partially mitigated this threat by replicating RQ1 and RQ2 on Python and JavaScript, using the same SOTorrent release, the same extraction pipeline, the same Siamese+ configuration and the same project popularity criteria~\cite{Ragkhitwetsagul2026}, as discussed in Section~\ref{subsec:replication}. Both findings hold in the two replication languages: accepted answers are edited even more frequently than in Java, and the number of matched answer edits again increases monotonically from low- to medium- to high-popularity projects.} 
Additionally, \rev{the 57 pull requests we opened, i.e., 24 carrying Fixing Bug and 33 carrying Improving Code recommendations, remain a small sample, and 37 of them are still awaiting a maintainer's review}. Thus, the findings from this pull request evaluation may not be representative of all the recommendations found in this study.
Finally, since this study focused on accepted answers, the findings may not be applicable to other types of SO answers (e.g., newest answers, highest-voted answers).
\rev{In addition, the SOTorrent dataset captures Stack Overflow activity only up to 31 December 2020, predating the widespread adoption of generative AI coding assistants. Editing behaviour and reuse dynamics on Stack Overflow may have shifted since then, so our findings reflect a historical snapshot and may not fully generalise to the current state of the platform.}

\section{Conclusions}
\label{sec:conclusion}

This paper has presented an empirical study of edits made to Stack Overflow Java accepted answers and used them to provide recommendations for open-source software projects. 
We analysed \rev{874,438} Java accepted answers on Stack Overflow and found that \rev{16.11\%} of them had been updated, i.e., contained more than one revision. The average number of revisions per accepted SO answer is 2.82, but it can reach 53.
By employing our specialised code clone search tool, called Siamese+, which can search Java code bases for similar code snippets across multiple revisions of Stack Overflow accepted answers and recommends code changes based on the latest code answer version, we found \rev{793 candidate code snippets in 10,673 GitHub Java projects, of which 391 (49.31\%) were manually judged applicable to the projects}. Our manual validation indicates that the latest code answers on Stack Overflow primarily fall into two categories: bug fixes and code improvements, found mostly in high-popularity GitHub projects. \rev{Measuring the same edits objectively with JavaParser and Checkstyle, we found that no structural or readability measure separates the latest snippets from the originals, since these edits are localised and semantic rather than restructuring, although they do tend to add defensive and error-handling logic.} \rev{We selected 107 recommendations for submission, of which 80 could still be applied to their target projects, and submitted them as 57 pull requests to the corresponding GitHub projects, i.e., 24 in the Fixing Bug category and 33 in the Improving Code category. At the time of writing, 12 pull requests comprising 19 code recommendations had been accepted by the project maintainers, i.e., 23.75\% of the 80 recommendations submitted, whereas 37 remained pending review and only five were closed without being adopted. Although the sample is small and the outcome of the pending pull requests is not yet known, these results provide preliminary evidence that such recommendations can be useful in real-world software projects.}

\rev{Lastly, a companion replication of RQ1 and RQ2 on Python and JavaScript indicates that both findings extend beyond Java~\cite{Ragkhitwetsagul2026}. In those two languages, 41.25\% and 39.10\% of accepted answers have been edited at least once, against 16.11\% for Java, while the number of revisions per edited answer is almost unchanged at 2.78 and 2.68 against 2.82, and the matched answer edits again concentrate in the most popular projects. The supply of potential code improvements is therefore larger outside Java than within it, although the replication covers neither the manual classification nor the pull-request evaluation, so the latter two findings remain specific to Java.}

The results from this study \rev{suggest} that Stack Overflow is a \rev{promising} source for creating Java code improvement recommendations based on its latest answer edits. Due to Stack Overflow's community-driven nature, answers are edited and improved over time, and these edits can be used to improve suboptimal code in software projects, whether the code was created independently or copied from Stack Overflow. The approach we used to locate potential clones between Stack Overflow and Java projects, i.e., Siamese+, \rev{provides a basis for} future studies on automated code recommendation tools. We foresee this automated tool being integrated into the code review process, where each new code update is checked against existing Stack Overflow answers and recommended updates are provided if needed.

\section{Declarations}



\subsection{Author contributions}
\begin{itemize}
    \item In-on Wiratsin: Conceptualization, Data Curation, Formal analysis, Investigation, Methodology, Software, Validation, Visualization, Writing - Original Draft, Writing - Review \& Editing
    \item Chaiyong Ragkhitwetsagul: Conceptualization, Data Curation, Formal analysis, Funding acquisition, Investigation, Methodology, Project administration, Resources, Software, Supervision, Validation, Visualization, Writing - Original Draft, Writing - Review \& Editing
    \item Matheus Paixao: Conceptualization, Formal analysis, Investigation, Methodology, Supervision, Validation, Writing - Original Draft, Writing - Review \& Editing
    \item Denis De Sousa: Conceptualization, Data Curation, Formal analysis, Investigation, Methodology, Software, Validation, Visualization, Writing - Original Draft, Writing - Review \& Editing
    \item Pongpop Lapvikai: Investigation, Methodology, Validation
    \item Peter Haddawy: Conceptualization, Methodology, Writing - Review \& Editing
\end{itemize}

\subsection{Data availability statement}
The authors declare that the data supporting the findings of this study are available within the paper and its Supporting Information files. \rev{The replication package, containing the datasets, the \textsc{Siamese+} source code, and the analysis scripts used in this study, is publicly archived at~\url{https://doi.org/10.5281/zenodo.17220745}.}

\subsection{Conflict of interest}
The authors have declared that no competing interests exist.

\section*{Supporting information}

\paragraph*{S1 Appendix.}
\label{S1_File}
{\bf Additional pull request examples.} Further examples of the pull requests
submitted to the GitHub projects, including the full maintainer discussion of the
bcgit pull request and two additional bug-fix recommendations.

%
%

\bibliography{references}

\begin{thebibliography}{10}

\bibitem{Shrestha2020}
Shrestha N, Botta C, Barik T, Parnin C.
\newblock Here We Go Again: Why is It Difficult for Developers to Learn Another
  Programming Language?
\newblock In: Proceedings of the ACM/IEEE 42nd International Conference on
  Software Engineering (ICSE '20); 2020. p. 691–701.

\bibitem{Eyolfson2011}
Eyolfson J, Tan L, Lam P.
\newblock {Do time of day and developer experience affect commit bugginess}.
\newblock In: Proceedings of the 8th Working Conference on Mining Software
  Repositories (MSR '11); 2011. p. 153.

\bibitem{Kuutila2020}
Kuutila M, M{\"{a}}ntyl{\"{a}} M, Farooq U, Claes M.
\newblock {Time pressure in software engineering: A systematic review}.
\newblock Information and Software Technology. 2020 May;121:106257.

\bibitem{Lavallee2015}
Lavallee M, Robillard PN.
\newblock {Why Good Developers Write Bad Code: An Observational Case Study of
  the Impacts of Organizational Factors on Software Quality}.
\newblock In: Proceedings of the 37th International Conference on Software
  Engineering (ICSE '15). vol.~1; 2015. p. 677-87.

\bibitem{Freire2020}
Freire S, Rios N, Gutierrez B, Torres D, Mendon{\c{c}}a M, Izurieta C, et~al.
\newblock {Surveying Software Practitioners on Technical Debt Payment Practices
  and Reasons for not Paying off Debt Items}.
\newblock In: Proceedings of the Evaluation and Assessment in Software
  Engineering (EASE '20); 2020. p. 210-9.

\bibitem{Graziotin2017}
Graziotin D, Fagerholm F, Wang X, Abrahamsson P.
\newblock {Unhappy Developers: Bad for Themselves, Bad for Process, and Bad for
  Software Product}.
\newblock In: Proceedings of the 39th International Conference on Software
  Engineering Companion (ICSE-C '17); 2017. p. 362-4.

\bibitem{Sjoberg2013}
Sjoberg DIK, Yamashita A, Anda BCD, Mockus A, Dyba T.
\newblock {Quantifying the Effect of Code Smells on Maintenance Effort}.
\newblock IEEE Transactions on Software Engineering. 2013 aug;39(8):1144-56.

\bibitem{Tom2013}
Tom E, Aurum A, Vidgen R.
\newblock {An exploration of technical debt}.
\newblock Journal of Systems and Software. 2013 Jun;86(6):1498-516.

\bibitem{Bavota2016}
Bavota G, Russo B.
\newblock {A large-scale empirical study on self-admitted technical debt}.
\newblock In: Proceedings of the 13th International Conference on Mining
  Software Repositories (MSR '16); 2016. p. 315-26.

\bibitem{Yang2017}
Yang D, Martins P, Saini V, Lopes C.
\newblock {Stack Overflow in GitHub:} Any Snippets There?
\newblock In: Proceedings of the 14th International Conference on Mining
  Software Repositories (MSR '17); 2017. p. 280-90.

\bibitem{Ragkhitwetsagul2019}
Ragkhitwetsagul C, Krinke J.
\newblock {Siamese: scalable and incremental code clone search via multiple
  code representations}.
\newblock Empirical Software Engineering. 2019 8;24:2236-84.

\bibitem{Zhang2019}
Zhang H, Wang S, hsun Peter~Chen T, Zou Y, Hassan AE.
\newblock {An Empirical Study of Obsolete Answers on Stack Overflow}.
\newblock IEEE Transactions on Software Engineering. 2019:850-62.

\bibitem{Chen2024}
Chen J, Zhao Y, Meng N.
\newblock How Do Developers Reuse StackOverflow Answers in Their GitHub
  Projects?
\newblock In: Proceedings of the 39th IEEE/ACM International Conference on
  Automated Software Engineering Workshops. New York, NY, USA: Association for
  Computing Machinery; 2024. p. 146–155.
\newblock \href {http://dx.doi.org/10.1145/3691621.3694945}
  {doi:10.1145/3691621.3694945}.

\bibitem{Huang2022}
Huang Y, Xu F, Zhou H, Chen X, Zhou X, Wang T.
\newblock Towards exploring the code reuse from stack overflow during software
  development.
\newblock In: Proceedings of the 30th IEEE/ACM International Conference on
  Program Comprehension. vol. 2022-March. ACM; 2022. p. 548-59.

\bibitem{Chen2024b}
Chen X, Xu F, Huang Y, Zhou X, Zheng Z.
\newblock {An empirical study of code reuse between GitHub and Stack Overflow
  during software development}.
\newblock Journal of Systems and Software. 2024;210:111964.
\newblock \href {http://dx.doi.org/https://doi.org/10.1016/j.jss.2024.111964}
  {doi:https://doi.org/10.1016/j.jss.2024.111964}.

\bibitem{Abdalkareem2025}
Abdalkareem R. {Studying the Role of Reusing Crowdsourcing Knowledge in
  Software Development}; 2025.
\newblock Available from: \url{https://arxiv.org/abs/2512.07824}. \href
  {http://arxiv.org/abs/2512.07824} {arXiv:2512.07824}.

\bibitem{Yu2026}
Yu C, Noei S, Zhang H, Zou Y.
\newblock An Empirical Study on the Characteristics of Reusable Code Clones.
\newblock ACM Trans Softw Eng Methodol. 2026 Jan.
\newblock Available from: \url{https://doi.org/10.1145/3793251}. \href
  {http://dx.doi.org/10.1145/3793251} {doi:10.1145/3793251}.

\bibitem{Gordon2024}
Burtch G, Lee D, Chen Z.
\newblock The consequences of generative AI for online knowledge communities.
\newblock Scientific Reports. 2024;14.
\newblock \href {http://dx.doi.org/https://doi.org/10.1038/s41598-024-61221-0}
  {doi:https://doi.org/10.1038/s41598-024-61221-0}.

\bibitem{delRioChanona2024}
del Rio-Chanona RM, Laurentsyeva N, Wachs J.
\newblock {Large language models reduce public knowledge sharing on online Q\&A
  platforms}.
\newblock PNAS Nexus. 2024 09;3(9):pgae400.
\newblock \href {http://dx.doi.org/10.1093/pnasnexus/pgae400}
  {doi:10.1093/pnasnexus/pgae400}.

\bibitem{Xue2026}
Xue J, Wang L, Zheng J, Li Y, Tan Y.
\newblock {Can ChatGPT Kill User-Generated Q\&A Platforms?}
\newblock Information Systems Research. 2026.
\newblock Articles in Advance.
\newblock \href {http://dx.doi.org/10.1287/isre.2023.0561}
  {doi:10.1287/isre.2023.0561}.

\bibitem{Shan2025}
Shan G, Qiu L.
\newblock {Examining the Impact of Generative AI on Users' Voluntary Knowledge
  Contribution: Evidence from a Natural Experiment on Stack Overflow}.
\newblock Information Systems Research. 2025.
\newblock \href {http://dx.doi.org/10.1287/isre.2023.0332}
  {doi:10.1287/isre.2023.0332}.

\bibitem{Helic2025}
Helic D, Santos T.
\newblock Stack Overflow is not dead yet: Crowd answers still matter.
\newblock Journal of Systems and Software. 2026;241:112988.
\newblock \href {http://dx.doi.org/https://doi.org/10.1016/j.jss.2026.112988}
  {doi:https://doi.org/10.1016/j.jss.2026.112988}.

\bibitem{Acar2016}
Acar Y, Backes M, Fahl S, Kim D, Mazurek ML, Stransky C.
\newblock You Get Where You're Looking for: The Impact of Information Sources
  on Code Security.
\newblock In: IEEE Symposium on Security and Privacy (SP '16); 2016. p.
  289-305.

\bibitem{Zhang2018}
Zhang T, Upadhyaya G, Reinhardt A, Rajan H, Kim M.
\newblock Are code examples on an online Q\&A forum reliable?
\newblock In: Proceedings of the 40th International Conference on Software
  Engineering (ICSE '18); 2018. p. 886-96.

\bibitem{Diamantopoulos2019}
Diamantopoulos T, Sifaki MI, Symeonidis A.
\newblock {Towards Mining Answer Edits to Extract Evolution Patterns in Stack
  Overflow}.
\newblock In: Proceedings of the 16th International Conference on Mining
  Software Repositories (MSR '19). vol. 2019-May; 2019. p. 216-9.

\bibitem{Tang2021}
Tang H, Nadi S.
\newblock {On using Stack Overflow comment-edit pairs to recommend code
  maintenance changes}.
\newblock Empirical Software Engineering. 2021 Jul;26(4):68.

\bibitem{Sheikhaei2023}
Sheikhaei MS, Tian Y, Wang S.
\newblock A study of update request comments in Stack Overflow answer posts.
\newblock Journal of Systems and Software. 2023 4;198:111590.

\bibitem{Mondal2025}
Mondal S, Roy CK.
\newblock {Does Editing Improve Answer Quality on Stack Overflow? A Data-Driven
  Investigation}.
\newblock In: Proceedings of the 41st International Conference on Software
  Maintenance and Evolution (ICSME'25); 2025. p. 492-504.
\newblock \href {http://dx.doi.org/10.1109/ICSME64153.2025.00051}
  {doi:10.1109/ICSME64153.2025.00051}.

\bibitem{Zuo2025}
Zuo S, Yang Q, Gao T, Progga ZT, Shen J.
\newblock How Security Coding Knowledge Impact Software Quality: An Empirical
  Study of Stack Overflow Security Issues.
\newblock In: 2025 25th International Conference on Software Quality,
  Reliability and Security (QRS); 2025. p. 727-38.
\newblock \href {http://dx.doi.org/10.1109/QRS65678.2025.00076}
  {doi:10.1109/QRS65678.2025.00076}.

\bibitem{Jallow2025}
Jallow A, Bugiel S.
\newblock {Stack Overflow Meets Replication: Security Research Amid Evolving
  Code Snippets}.
\newblock In: Proceedings of the 34th USENIX Security Symposium (USENIX
  Security'25). USENIX Association; 2025. p. 2809-28.
\newblock Extended version: arXiv:2501.16948.

\bibitem{Keivanloo2014}
Keivanloo I, Rilling J, Zou Y.
\newblock Spotting working code examples.
\newblock In: Proceedings of the 36th International Conference on Software
  Engineering (ICSE '14); 2014. p. 664-75.

\bibitem{Ponzanelli2013}
Ponzanelli L, Bacchelli A, Lanza M.
\newblock {Seahawk: Stack Overflow in the IDE}.
\newblock In: Proceedings of the 2013 International Conference on Software
  Engineering (ICSE '13); 2013. p. 1295-8.

\bibitem{Ponzanelli2014}
Ponzanelli L, Bavota G, {Di Penta} M, Oliveto R, Lanza M.
\newblock {Mining StackOverflow to turn the IDE into a self-confident
  programming prompter}.
\newblock In: Proceedings of the 14th International Conference on Mining
  Software Repositories (MSR '14); 2014. p. 102-11.

\bibitem{Treude2016}
Treude C, Robillard MP.
\newblock {Augmenting API documentation with insights from stack overflow}.
\newblock In: Proceedings of the 38th International Conference on Software
  Engineering (ICSE '16). ACM Press; 2016. p. 392-403.

\bibitem{Gao2023}
Gao Z, Xia X, Lo D, Grundy J, Zhang X, Xing Z.
\newblock I Know What You Are Searching for: Code Snippet Recommendation from
  Stack Overflow Posts.
\newblock ACM Transactions on Software Engineering and Methodology. 2023
  7;32:1-42.
\newblock \href {http://dx.doi.org/10.1145/3550150} {doi:10.1145/3550150}.

\bibitem{Jallow2024}
Jallow A, Schilling M, Backes M, Bugiel S.
\newblock Measuring the Effects of Stack Overflow Code Snippet Evolution on
  Open-Source Software Security.
\newblock In: Proceedings of the 2024 IEEE Symposium on Security and Privacy
  (SP'24). IEEE; 2024. p. 1083-101.
\newblock \href {http://dx.doi.org/10.1109/SP54263.2024.00022}
  {doi:10.1109/SP54263.2024.00022}.

\bibitem{Wang2020}
Wang S, Chen TH, Hassan AE.
\newblock How Do Users Revise Answers on Technical Q\&A Websites? A Case Study
  on Stack Overflow.
\newblock IEEE Transactions on Software Engineering. 2020 9;46:1024-38.
\newblock \href {http://dx.doi.org/10.1109/TSE.2018.2874470}
  {doi:10.1109/TSE.2018.2874470}.

\bibitem{Baltes2020}
Baltes S, Wagner M.
\newblock {An annotated dataset of stack overflow post edits}.
\newblock In: Proceedings of the 2020 Genetic and Evolutionary Computation
  Conference Companion (GECCO '20). vol. 323. ACM; 2020. p. 1923-5.

\bibitem{Zhang2022}
Zhang H, Wang S, Li H, Chen TH, Hassan AE.
\newblock {A Study of C/C++ Code Weaknesses on Stack Overflow}.
\newblock IEEE Transactions on Software Engineering. 2022 7;48:2359-75.
\newblock \href {http://dx.doi.org/10.1109/TSE.2021.3058985}
  {doi:10.1109/TSE.2021.3058985}.

\bibitem{Rahman2016}
Rahman MM, Roy CK, Lo D.
\newblock {RACK: Automatic API Recommendation Using Crowdsourced Knowledge}.
\newblock In: Proceedings of the 23rd IEEE International Conference on Software
  Analysis, Evolution, and Reengineering (SANER '16); 2016. p. 349-59.

\bibitem{Rubei2020}
Rubei R, {Di Sipio} C, Nguyen PT, {Di Rocco} J, {Di Ruscio} D.
\newblock {PostFinder: Mining Stack Overflow posts to support software
  developers}.
\newblock Information and Software Technology. 2020 Nov;127:106367.

\bibitem{Nyamawe2018}
Nyamawe AS, Liu H, Niu Z, Wang W, Niu N.
\newblock {Recommending Refactoring Solutions Based on Traceability and Code
  Metrics}.
\newblock IEEE Access. 2018;6:49460-75.

\bibitem{Jin2023}
Jin Y, Bai Y, Zhu Y, Sun Y, Wang W.
\newblock Code Recommendation for Open Source Software Developers.
\newblock In: Proceedings of the ACM Web Conference 2023. ACM; 2023. p.
  1324-33.
\newblock \href {http://dx.doi.org/10.1145/3543507.3583503}
  {doi:10.1145/3543507.3583503}.

\bibitem{Luan2019}
Luan S, Yang D, Barnaby C, Sen K, Chandra S.
\newblock Aroma: Code Recommendation via Structural Code Search.
\newblock Proceedings of the ACM on Programming Languages.
  2019;3(OOPSLA):1–-28.

\bibitem{Guo2024}
Guo Q, Cao J, Xie X, Liu S, Li X, Chen B, et~al.
\newblock Exploring the Potential of ChatGPT in Automated Code Refinement: An
  Empirical Study.
\newblock In: Proceedings of the IEEE/ACM 46th International Conference on
  Software Engineering. ACM; 2024. p. 1-13.

\bibitem{Baltes2018a}
Baltes S, Dumani L, Treude C, Diehl S.
\newblock {SOTorrent:} Reconstructing and Analyzing the Evolution of {Stack
  Overflow} Posts.
\newblock In: Proceedings of the 15th International Conference on Mining
  Software Repositories (MSR '18); 2018. p. 319-30.

\bibitem{Manes2019}
Manes SS, Baysal O.
\newblock How Often and What StackOverflow Posts Do Developers Reference in
  Their GitHub Projects?
\newblock In: 2019 IEEE/ACM 16th International Conference on Mining Software
  Repositories (MSR); 2019. p. 235-9.

\bibitem{Bangash2019}
Bangash AA, Sahar H, Chowdhury S, Wong AW, Hindle A, Ali K.
\newblock What do Developers Know About Machine Learning: A Study of ML
  Discussions on StackOverflow.
\newblock In: 2019 IEEE/ACM 16th International Conference on Mining Software
  Repositories (MSR); 2019. p. 260-4.

\bibitem{Nugroho2022}
Nugroho YS, Islam S, Gunawan D, Kurniawan YI, Hossain MJ.
\newblock Dataset of network simulator related-question posts in stack
  overflow.
\newblock Data in Brief. 2022;41:107942.
\newblock \href {http://dx.doi.org/https://doi.org/10.1016/j.dib.2022.107942}
  {doi:https://doi.org/10.1016/j.dib.2022.107942}.

\bibitem{Dabic:msr2021data}
Dabic O, Aghajani E, Bavota G.
\newblock Sampling Projects in GitHub for {MSR} Studies.
\newblock In: Proceedings of the 18th International Conference on Mining
  Software Repositories (MSR '21); 2021. p. 560-4.

\bibitem{IEEETopProg2024}
Cass S. The Top Programming Languages 2024; 2024.
\newblock \url{https://spectrum.ieee.org/top-programming-languages-2024}.

\bibitem{Zakeri2023}
Zakeri-Nasrabadi M, Parsa S, Ramezani M, Roy C, Ekhtiarzadeh M.
\newblock A systematic literature review on source code similarity measurement
  and clone detection: Techniques, applications, and challenges.
\newblock Journal of Systems and Software. 2023 10;204:111796.
\newblock \href {http://dx.doi.org/10.1016/j.jss.2023.111796}
  {doi:10.1016/j.jss.2023.111796}.

\bibitem{Sajnani2016}
Sajnani H, Saini V, Svajlenko J, Roy CK, Lopes CV.
\newblock {SourcererCC: Scaling Code Clone Detection to Big-Code}.
\newblock In: Proceedings of the 38th International Conference on Software
  Engineering (ICSE '16). ACM Press; 2016. p. 1157-68.

\bibitem{Ragkhitwetsagul2018}
Ragkhitwetsagul C, Krinke J, Clark D.
\newblock {A comparison of code similarity analysers}.
\newblock Empirical Software Engineering. 2018 August;23(4):2464-519.

\bibitem{Ragkhitwetsagul2019a}
Ragkhitwetsagul C, Krinke J, Paixao M, Bianco G, Oliveto R.
\newblock Toxic Code Snippets on Stack Overflow.
\newblock IEEE Transactions on Software Engineering. 2021;47(3):560-81.

\bibitem{Ragkhitwetsagul2021}
Ragkhitwetsagul C, Krinke J, Paixao M, Bianco G, Oliveto R.
\newblock Toxic Code Snippets on Stack Overflow.
\newblock IEEE Transactions on Software Engineering. 2021;47(3):560-81.
\newblock \href {http://dx.doi.org/10.1109/TSE.2019.2900307}
  {doi:10.1109/TSE.2019.2900307}.

\bibitem{tempero2010qualitas}
Tempero E, Anslow C, Dietrich J, Han T, Li J, Lumpe M, et~al.
\newblock The qualitas corpus: A curated collection of java code for empirical
  studies.
\newblock In: 2010 Asia pacific software engineering conference. IEEE; 2010. p.
  336-45.

\bibitem{wang2013searching}
Wang T, Harman M, Jia Y, Krinke J.
\newblock Searching for better configurations: a rigorous approach to clone
  evaluation.
\newblock In: FSE '13; 2013. p. 455-65.

\bibitem{Ragkhitwetsagul2016}
Ragkhitwetsagul C, Paixao M, Adham M, Busari S, Krinke J, Drake JH.
\newblock Searching for configurations in clone evaluation--a replication
  study.
\newblock In: SSBSE '16; 2016. p. 250-6.

\bibitem{Zampetti2019}
Zampetti F, Bavota G, Canfora G, Penta MD.
\newblock {A Study on the Interplay between Pull Request Review and Continuous
  Integration Builds}.
\newblock In: Proceedings of the 26th IEEE International Conference on Software
  Analysis, Evolution and Reengineering (SANER '19); 2019. p. 38-48.

\bibitem{Gonzalez2020}
Gonzalez D, Zimmermann T, Nagappan N.
\newblock {The State of the ML-universe}.
\newblock In: Proceedings of the 17th International Conference on Mining
  Software Repositories (MSR '20). ACM; 2020. p. 431-42.

\bibitem{Sheoran2014}
Sheoran J, Blincoe K, Kalliamvakou E, Damian D, Ell J.
\newblock {Understanding "watchers" on GitHub}.
\newblock In: Proceedings of the 14th International Conference on Mining
  Software Repositories (MSR '14); 2014. p. 336-9.

\bibitem{Munaiah2017}
Munaiah N, Kroh S, Cabrey C, Nagappan M.
\newblock {Curating GitHub for engineered software projects}.
\newblock Empirical Software Engineering. 2017 Dec;22(6):3219-53.

\bibitem{javaparser}
{The JavaParser Team}. {JavaParser}: Analyse, Transform and Generate Your
  {Java} Code Base; 2025.
\newblock Version 3.28.1.
\newblock \url{https://javaparser.org/}.

\bibitem{javaparser-visited}
Smith N, van Bruggen D, Tomassetti F.
\newblock {JavaParser}: Visited.
\newblock Leanpub; 2021.
\newblock Reference guide to the {JavaParser} library.

\bibitem{checkstyle}
{Checkstyle Development Team}. {Checkstyle}: A Development Tool to Help
  Programmers Write {Java} Code That Adheres to a Coding Standard; 2025.
\newblock Version 13.9.0.
\newblock \url{https://checkstyle.org/}.

\bibitem{Roy2009}
Roy CK, Cordy JR, Koschke R.
\newblock {Comparison and evaluation of code clone detection techniques and
  tools: A qualitative approach}.
\newblock Science of Computer Programming. 2009;74(7):470-95.

\bibitem{shapiro1965}
Shapiro SS, Wilk MB.
\newblock An analysis of variance test for normality (complete samples).
\newblock Biometrika. 1965;52(3--4):591-611.

\bibitem{kruskal1952}
Kruskal WH, Wallis WA.
\newblock Use of ranks in one-criterion variance analysis.
\newblock Journal of the American Statistical Association.
  1952;47(260):583-621.

\bibitem{Williams2023}
Williams D, Callan J, Kirbas S, Mechtaev S, Petke J, Prideaux-Ghee T, et~al.
\newblock User-Centric Deployment of Automated Program Repair at Bloomberg.
\newblock In: Proceedings of the 46th International Conference on Software
  Engineering: Software Engineering in Practice. ICSE-SEIP '24. New York, NY,
  USA: Association for Computing Machinery; 2024. p. 81-91.
\newblock \href {http://dx.doi.org/10.1145/3639477.3639756}
  {doi:10.1145/3639477.3639756}.

\bibitem{Watanabe2025}
Watanabe M, Kula RG, Fujino S, Uchida S, Fowze F, Zhu H, et~al.
\newblock {On the Use of Agentic Coding: An Empirical Study of Pull Requests on
  GitHub}.
\newblock ACM Transactions on Software Engineering and Methodology. 2025.

\bibitem{Pinto2018}
Pinto G, Steinmacher I, Gerosa MA.
\newblock {Who Gets a Patch Accepted First? Comparing the Contributions of
  Employees and Volunteers}.
\newblock In: Proceedings of the 11th International Workshop on Cooperative and
  Human Aspects of Software Engineering (CHASE); 2018. p. 110-3.
\newblock \href {http://dx.doi.org/10.1145/3195836.3195858}
  {doi:10.1145/3195836.3195858}.

\bibitem{Ragkhitwetsagul2026}
Ragkhitwetsagul C, Wiratsin Io, Paixao M, De~Sousa D, Lapvikai P, Haddawy P.
  {Do Stack Overflow Answer Edits Occur Beyond Java? A Replication on Python
  and JavaScript}; 2026.
\newblock Available from: \url{https://arxiv.org/abs/2608.08361}. \href
  {http://arxiv.org/abs/2608.08361} {arXiv:2608.08361}.

\bibitem{Liu2024}
Liu Y, Le-Cong T, Widyasari R, Tantithamthavorn C, Li L, Le XBD, et~al.
\newblock Refining ChatGPT-Generated Code: Characterizing and Mitigating Code
  Quality Issues.
\newblock ACM Transactions on Software Engineering and Methodology. 2024
  6;33:1-26.
\newblock \href {http://dx.doi.org/10.1145/3643674} {doi:10.1145/3643674}.

\bibitem{Wang2024}
Wang J, Luo X, Cao L, He H, Huang H, Xie J, et~al.. Is Your AI-Generated Code
  Really Safe? Evaluating Large Language Models on Secure Code Generation with
  CodeSecEval; 2024.
\newblock Available from: \url{https://arxiv.org/abs/2407.02395}. \href
  {http://arxiv.org/abs/2407.02395} {arXiv:2407.02395}.

\bibitem{Fu2025}
Fu Y, Liang P, Tahir A, Li Z, Shahin M, Yu J, et~al.
\newblock Security Weaknesses of Copilot-Generated Code in GitHub Projects: An
  Empirical Study.
\newblock ACM Trans Softw Eng Methodol. 2025 Oct;34(8).
\newblock \href {http://dx.doi.org/10.1145/3716848} {doi:10.1145/3716848}.

\bibitem{Siddiq2024}
Siddiq ML, Roney L, Zhang J, Santos JCDS.
\newblock Quality Assessment of ChatGPT Generated Code and their Use by
  Developers.
\newblock In: {Proceedings of the 21st International Conference on Mining
  Software Repositories (MSR '24)}. New York, NY, USA; 2024. p. 152–156.
\newblock \href {http://dx.doi.org/10.1145/3643991.3645071}
  {doi:10.1145/3643991.3645071}.

\bibitem{Shihan2024}
Dou S, Jia H, Wu S, Zheng H, Zhou W, Wu M, et~al.. What's Wrong with Your Code
  Generated by Large Language Models? An Extensive Study; 2024.
\newblock Available from: \url{https://arxiv.org/abs/2407.06153}. \href
  {http://arxiv.org/abs/2407.06153} {arXiv:2407.06153}.

\bibitem{Tambon2025}
Tambon F, Moradi-Dakhel A, Nikanjam A, Khomh F, Desmarais MC, Antoniol G.
\newblock Bugs in large language models generated code: an empirical study.
\newblock Empirical Software Engineering. 2025 feb;30(3):65.
\newblock \href {http://dx.doi.org/10.1007/s10664-025-10614-4}
  {doi:10.1007/s10664-025-10614-4}.

\bibitem{Cotroneo2025}
Cotroneo D, Improta C, Liguori P. Human-Written vs. AI-Generated Code: A
  Large-Scale Study of Defects, Vulnerabilities, and Complexity; 2025.
\newblock Available from: \url{https://arxiv.org/abs/2508.21634}. \href
  {http://arxiv.org/abs/2508.21634} {arXiv:2508.21634}.

\bibitem{Verdi2020}
Verdi M, Sami A, Akhondali J, Khomh F, Uddin G, Motlagh AK.
\newblock An Empirical Study of C++ Vulnerabilities in Crowd-Sourced Code
  Examples.
\newblock IEEE Transactions on Software Engineering. 2022;48(5):1497-514.
\newblock \href {http://dx.doi.org/10.1109/TSE.2020.3023664}
  {doi:10.1109/TSE.2020.3023664}.

\end{thebibliography}

\clearpage
\appendix
\section*{S1 Appendix. Additional pull request examples.}
\setcounter{figure}{0}
\renewcommand{\figurename}{}
\renewcommand{\thefigure}{S\arabic{figure} Fig}

\bigskip
\noindent This appendix contains additional examples of the pull requests submitted to the
GitHub projects, referred to from the \textit{Practical application of the approach
(Phase 4)} section of the manuscript.

\bigskip

\begin{figure}[htbp]
    \centering
    \includegraphics[width=0.8\linewidth]{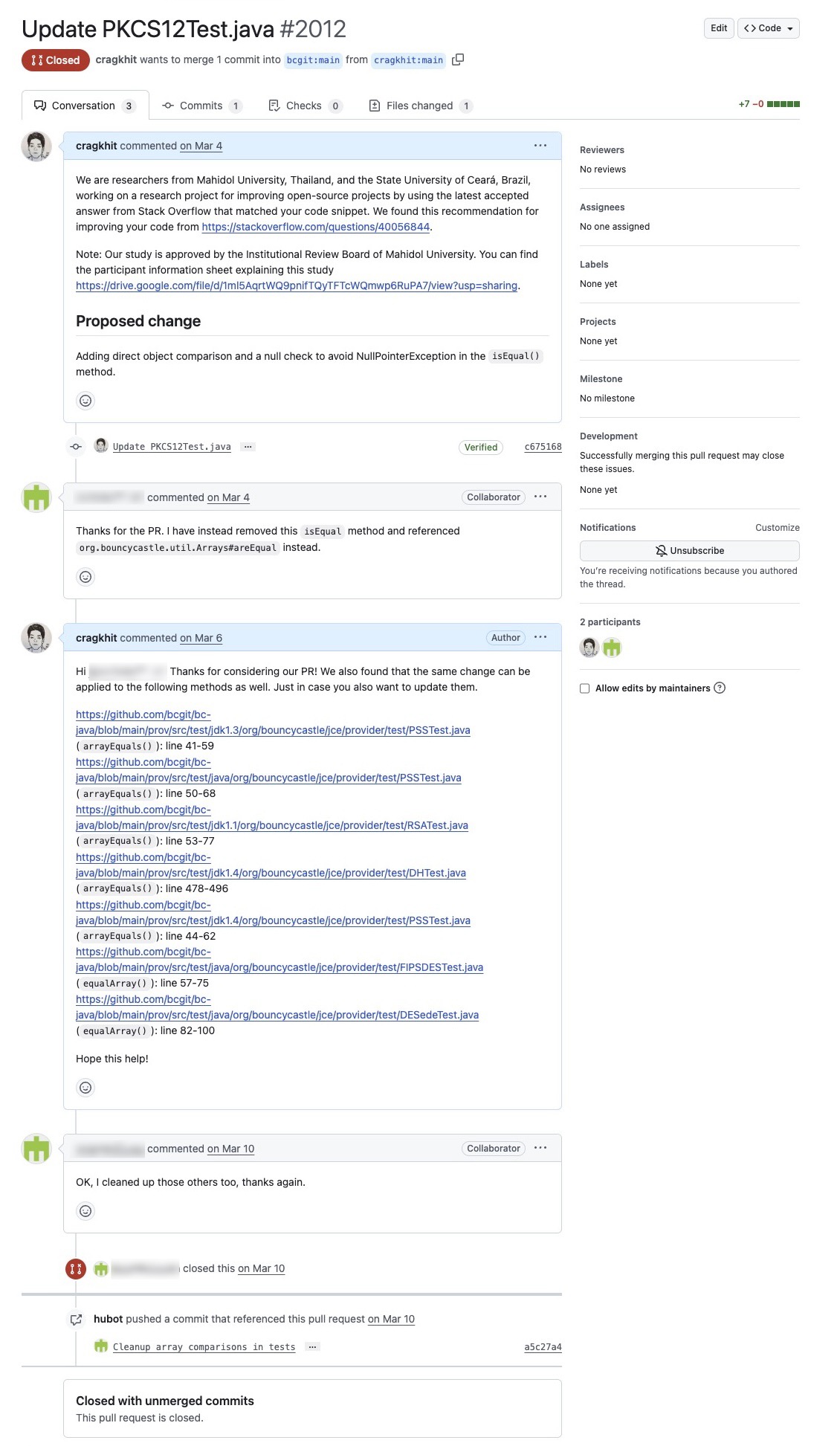}
    \caption{\rev{Pull request of the bcgit project: a change to \texttt{isEqual()} avoiding a NullPointerException, followed by seven similar changes in discussion with the maintainer.}}
    \label{fig:pr_bcgit}
\end{figure}

\begin{figure}[htbp]
\begin{adjustwidth}{-2.25in}{0in}
    \begin{lstlisting}[language=diff]
@@ -93,16 +93,21 @@ private final static String escapeEnding(String s) {
	// File utils
	public final static void copy(File src, File dst) throws IOException {
		InputStream in = new FileInputStream(src);
-       OutputStream out = new FileOutputStream(dst);
-       
-       // Transfer bytes from in to out
-       byte[] buf = new byte[1024];
-       int len;
-       while ((len = in.read(buf)) > 0) {
-           out.write(buf, 0, len);
-       }
-       in.close();
-       out.close();
+           try {
+               OutputStream out = new FileOutputStream(dst);
+               try {
+            			// Transfer bytes from in to out
+            			byte[] buf = new byte[1024];
+            			int len;
+            			while ((len = in.read(buf)) > 0) {
+                			out.write(buf, 0, len);
+            			}
+        		} finally {
+            			out.close();
+        		}
+    		} finally {
+        		in.close();
+    		}
	}
    \end{lstlisting}
    \caption{\rev{Bug-fix recommendation in the cyriux project pull request, ensuring proper closing of the OutputStream \texttt{out} (\url{https://github.com/cyriux/mpcmaid/pull/16}).}}
\end{adjustwidth}
\end{figure}

\begin{figure}[htbp]
\begin{adjustwidth}{-2.25in}{0in}
    \begin{lstlisting}[language=diff]
@@ -6,6 +6,9 @@
public class AddSubtractDaysSkippingWeekendsUtils {

    public static LocalDate addDaysSkippingWeekends(LocalDate date, int days) {
+       if (days < 1) {
+           return date;
+       }
        LocalDate result = date;
        int addedDays = 0;
        while (addedDays < days) {
@@ -18,6 +21,9 @@ public static LocalDate addDaysSkippingWeekends(LocalDate date, int days) {
    }

    public static LocalDate subtractDaysSkippingWeekends(LocalDate date, int days) {
+       if (days < 1) {
+           return date;
+       }
        LocalDate result = date;
        int subtractedDays = 0;
        while (subtractedDays < days) {
    \end{lstlisting}
    \caption{\rev{Bug-fix recommendation in the eugenp project pull request, checking the invalid \texttt{days} value (\url{https://github.com/eugenp/tutorials/pull/18370}).}}
\end{adjustwidth}
\end{figure}

\end{document}